\documentclass[aps,twocolumn,groupedaddress]{revtex4}

\usepackage{graphicx}  % needed for figures
\usepackage{dcolumn}   % needed for some tables
\usepackage{bm}        % for math
\usepackage{amssymb}   % for math
\usepackage{amsmath}
\usepackage{epsfig}
\usepackage{xspace}
\usepackage{hyperref}
\usepackage{color}
\usepackage[utf8]{inputenc}

\newcommand{\bea}{\begin{eqnarray}}
\newcommand{\eea}{\end{eqnarray}}

\definecolor{MyDarkBlue}{rgb}{0,0.0,0.7}
\hypersetup{pdfborder={0 0 0},colorlinks,breaklinks=true,
urlcolor={MyDarkBlue},citecolor={MyDarkBlue},linkcolor={MyDarkBlue}
}

\begin{document}

% the following line is for submission, including submission to the arXiv!!
%\hspace{5.2in} \mbox{Darmstadt/DESY/U-Tokyo}

\title{Deconfinement transition line with the Complex Langevin equation up to $\mu /T \sim 5$ }
\author{M. Scherzer$^{1}$}
\author{D. Sexty$^{2,3}$}
\author{I.-O. Stamatescu$^{1}$}

\affiliation{$^1$ {\it Institut f\"ur Theoretische Physik, Universit\"at Heidelberg, Philosophenweg 16, 69120 Heidelberg, Germany}}
\affiliation{ $^2$ {\it Department of Physics, Wuppertal University, Gaussstr. 20, D-42119 Wuppertal, Germany}}
\affiliation{ $^3$ {\it J\"ulich Supercomputing Centre, Forschungszentrum J\"ulich, D-52425 J\"ulich, Germany}}

\begin{abstract} 
  We study the deconfinement transition line in QCD for quark chemical
  potentials up to $\mu_q \sim 5 T$ ($\mu_B \sim 15 T$). To circumvent
  the sign problem we use the complex Langevin equation with gauge cooling.
  The plaquette gauge action is used with two flavors of naive Wilson
  fermions at a relatively heavy pion mass of roughly 1.3 GeV.
  A quadratic dependence describes the transition line well on the whole
  chemical potential range. 
\end{abstract}
%\pacs{11.15.Ha, 12.38.Gc}

\maketitle

\section{Introduction}

The study of the phase diagram of QCD on the Temperature($T$)-quark chemical
potential ($\mu$) plane using first principles
methods is hampered by the sign problem at $\mu>0$, which
invalidates naive Monte-Carlo simulations using importance sampling.
The $\mu=0$ axis is well known
\cite{Petreczky:2012rq,Philipsen:2012nu,Borsanyi:2016bzg}, which
features a crossover phase transition
around $ T \approx 150$ MeV for physical quark masses. The common lore
suggests that this phase transition should get stronger
as the chemical potential is increased, eventually
reaching a critical point and changing into a first order phase
transition line for higher $\mu$.

In order to investigate the transition line in QCD, one has to circumvent
the sign problem in some way. Previous studies used the
reweighting method \cite{Barbour:1997ej,Fodor:2001pe,DePietri:2007ak}, the Taylor 
expansion from $\mu =0$ \cite{deForcrand:1999ih,Miyamura:2002mpl,Kaczmarek:2011zz,Endrodi:2011gv,Bonati:2018nut,Bazavov:2018mes} or analytic continuation from $\mu \leq 0$
\cite{Cea:2014xva,Bonati:2015bha,Bellwied:2015rza,Borsanyi:2020fev}. 
These methods deliver solid results for quark
chemical potentials up to $ \mu_q/T \simeq 1 $.

In the literature the small chemical potential behavior
of the transition line is approximated with a polynomial behavior
(using the baryon chemical potential $\mu_B=3 \mu_q$):
\bea \label{mubarionkappa}
{ T_c(\mu_B) \over T_c(0) } = 1 - \kappa_2 \left({\mu_B \over T_c(0) }\right)^2 - \kappa_4 \left( {\mu_B\over T_c(0)}\right)^4 + O(\mu_B^6).
\eea
(In the following $\mu$ denotes the quark chemical potential.)
The value of the curvature $\kappa_2$ turns out to be quite
small \cite{Endrodi:2011gv,Bonati:2018nut}, just as $\kappa_4$, which
is consistent with zero within
the statistical errorbars of the state of the art \cite{Borsanyi:2020fev}.

In this paper we
employ the Complex Langevin (CL)
equation \cite{Parisi:1984cs,Klauder:1983nn}.
In the last decade the method has enjoyed a revival of interest, initiated
by studies aiming at application for physical simulations \cite{Berges2005b,Aarts:2008rr}
(see also the recent reviews \cite{Berger:2019odf,Attanasio:2020spv}).
The CL equation complexifies the field manifold using analytical continuation
of the variables (not to be confused with the analytical continuation
in $\mu$ mentioned above)
to circumvent the sign problem, and allows for direct simulations
at $\mu>0$.
The aim of this study is to show that the CL simulations
allow following the transition line to previously
unaccessible chemical potential values.
Here we use the plaquette gauge action with naive Wilson
fermions with relatively heavy quark masses to study the transition
line for $\mu / T$ values up to 5.

In Sec.~\ref{secsetup} we describe the setup of our simulations
and discuss
%%%%%N  it's
their
%%%%%N
 behavior as we get closer to the continuum limit as well
as other known issues of CL simulations. 
In Sec.~\ref{secmethods} we present in detail our methods for mapping
out the phase transition line, and also present the numerical results.
Finally we conclude in Sec.~\ref{secconc}.

\section{Simulation setup}
\label{secsetup}
\subsection{Complex Langevin for QCD}
\label{secsetupCL}
For the link variables
$U_{x,\nu}$ of gauge theories on the SU(N) manifold the discretised 
update with Langevin timestep $\epsilon$  
is written as \cite{PhysRevD.32.2736}:

\bea
U_{x,\nu} (\tau+\epsilon) = 
  \textrm{exp} \left[ i \sum\limits_a 
 \lambda_a ( \epsilon K_{ax\nu}  + \sqrt\epsilon \eta_{ax\nu} ) 
\right]         U_{x,\nu}(\tau),
\eea
with $\lambda_a$ the generators of the gauge group, i.e. 
the Gell-Mann matrices, and a Gaussian white noise $\eta_{ax\nu}$ with
$ \langle \eta_{ax\alpha} \eta_{by\beta} \rangle = 2 \delta_{ab} \delta_{xy}\delta_{\alpha\beta} $.
The drift force $K_{ax\nu} = -D_{ax\nu} S[U]   $ is
calculated from the action using the left derivative
\bea
D_{ax\nu}f(U) = \left. \partial_\alpha f( e^{i \alpha \lambda_a} U_{x,\nu}) \right|_{\alpha=0}.
\eea
For complex actions the drift force $ K_{ax\nu}$ is in general complex,
thus the manifold of the link variables has to be complexified to SL(N,$\mathcal{C}$). 
In the case of lattice QCD with fermions the measure of the
theory is written as 
\bea
\rho_\textrm{eff} = e^{-S_\textrm{YM}} \textrm{det} M(\mu)
\eea
with the determinant of the fermionic Dirac matrix $M(\mu)$,
resulting in a non-holomorphic action
$S_{eff}= S_{YM} - \textrm{ln det} M(\mu)$, and meromorphic drift terms.
Simulating such a theory is not guaranteed to give correct results if 
the zeroes of the measure are visited  
by the process \cite{Mollgaard:2013qra,Greensite:2014cxa,Nishimura:2015pba,Aarts:2017vrv}.

\begin{figure}
\includegraphics[width=0.48\textwidth]{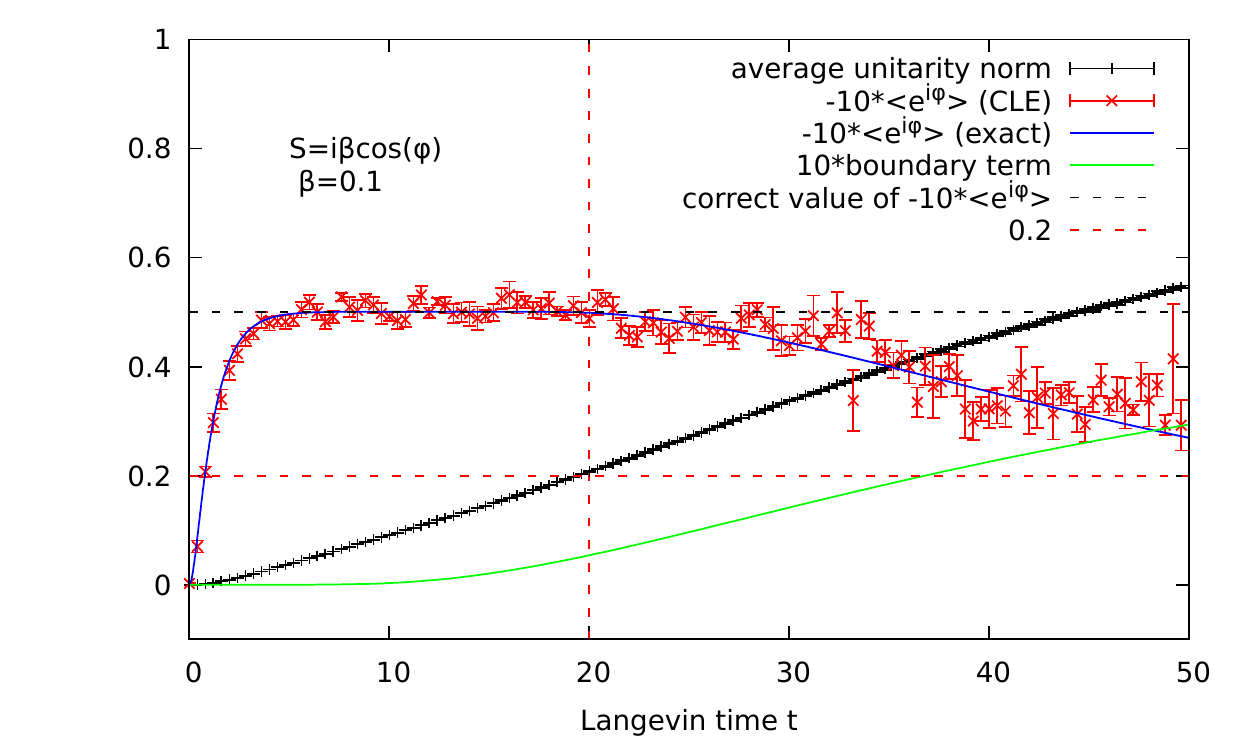}
\caption{Behavior of the observable $e^{i\varphi}$, the unitarity norm and the boundary term in a simple U(1) plaquette model
 with action $ S=  i \beta \cos(\varphi) $, with $\beta=0.1$ as investigated in \cite{boundaryterms1}.}
\label{fig:unorm_cutoff}
\end{figure}

To monitor the process on the complex manifold SL(3,$\mathcal{C}$), we use the unitarity norm (UN)
\bea
\label{UN}
N_U&=& {1 \over 4 \Omega} \sum_{x,\nu} \sum_{i,k} \left| ( U_{x,\nu} U^\dagger_{x,\nu} - 1)  \right|^2_{ik}.
\eea
where $\Omega = N^3_s N_t $ is the space-time volume of the lattice.
To avoid an uncontrolled growth of the unitarity norm and thus a quick
breakdown of the simulation we use the gauge cooling
procedure \cite{gaugecooling,Aarts:2013uxa}. The system then shows a
quick thermalization of the physical quantities such as the plaquette or Polyakov loop (well before Langevin time $\tau=10$ is reached), while the
unitarity norm tends to grow slowly, and saturates for large
Langevin time $\tau$ around $N_U \sim O(1)$. 
As observed earlier \cite{hdqcdpd}, in this state,
in spite of the gauge cooling, the process 
has distributions with slow decay in the non-compact directions
where the boundary terms
can no longer be neglected and spoil the correctness proof of the method \cite{boundaryterms1}. We therefore use the first part of
the Langevin time evolution after the physical quantities have equilibrated but
the unitarity norm is still small.
%\subsection{Why we can cut the simulation before thermalization of the unitarity norm}
This is motivated by the observation that oftentimes when CL simulations yield wrong results they do so only after a certain time, i.e.~they first thermalize to the correct solution and after some time start deviating from this solution again. This has been observed in a simple U(1) plaquette model in \cite{boundaryterms1}. We show the behavior of the unitarity norm in this model, the observable $e^{i\varphi}$ as well as the boundary term as a function of Langevin time in Fig.~\ref{fig:unorm_cutoff}. Here, the observable first thermalizes to the correct value and stays at this value up to $t\approx 20$. At $t\approx 20$ the process develops a non-negligible boundary term, which happens when the unitarity norm reached a value above $UN=0.2$. Hence, in this model we would retain correct expectation values if we stopped the simulation at an average unitarity norm of that value. 
Similar behavior has been observed in real time simulations for SU(2) gauge theory in \cite{Berges:2006xc,Berges:2007nr} for the spatial plaquette.
The generalization to QCD from those simple examples is not straight forward. Hence, we first look at HDQCD \cite{Bender:1992gn,Blum:1995cb}, where
the spatial hopping terms in the Dirac matrix are dropped and it is
a good approximation to full QCD for heavy quarks at high density. In HDQCD CL works very well in certain parameter regions and it was possible to map out the full phase diagram \cite{hdqcdpd}.

\begin{figure}
\includegraphics[width=0.48\textwidth]{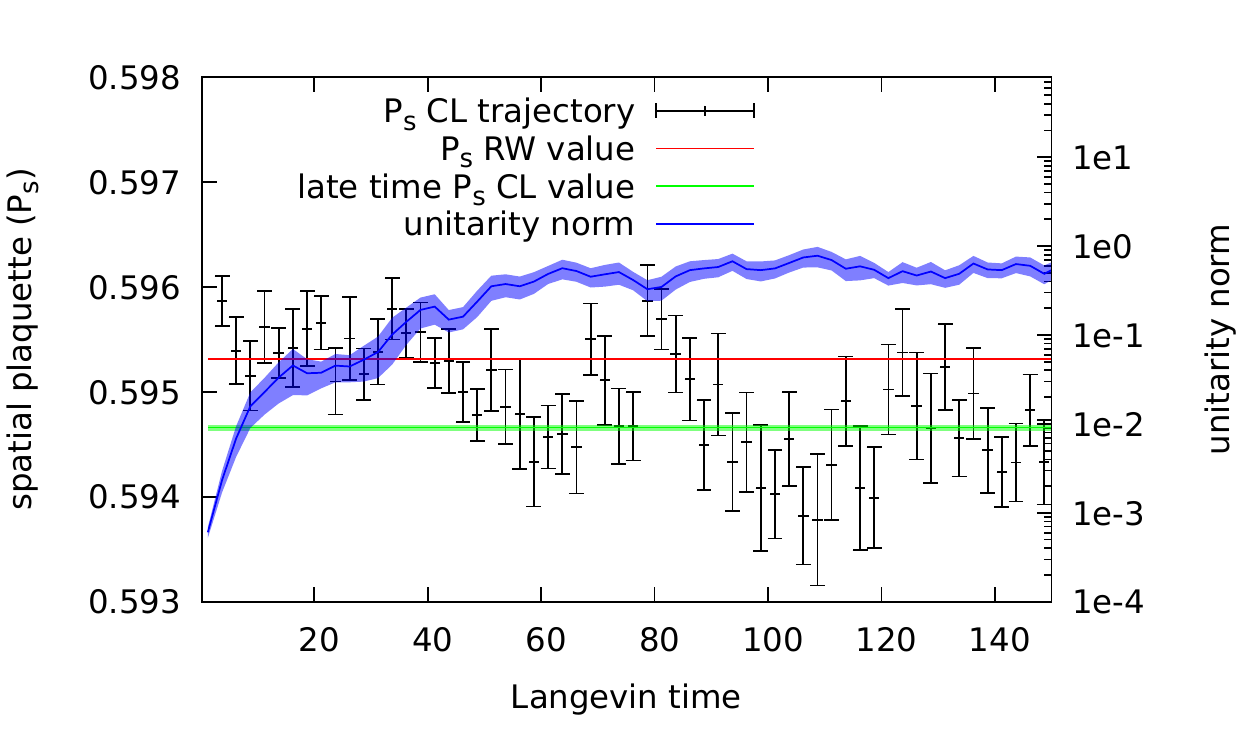}
\includegraphics[width=0.48\textwidth]{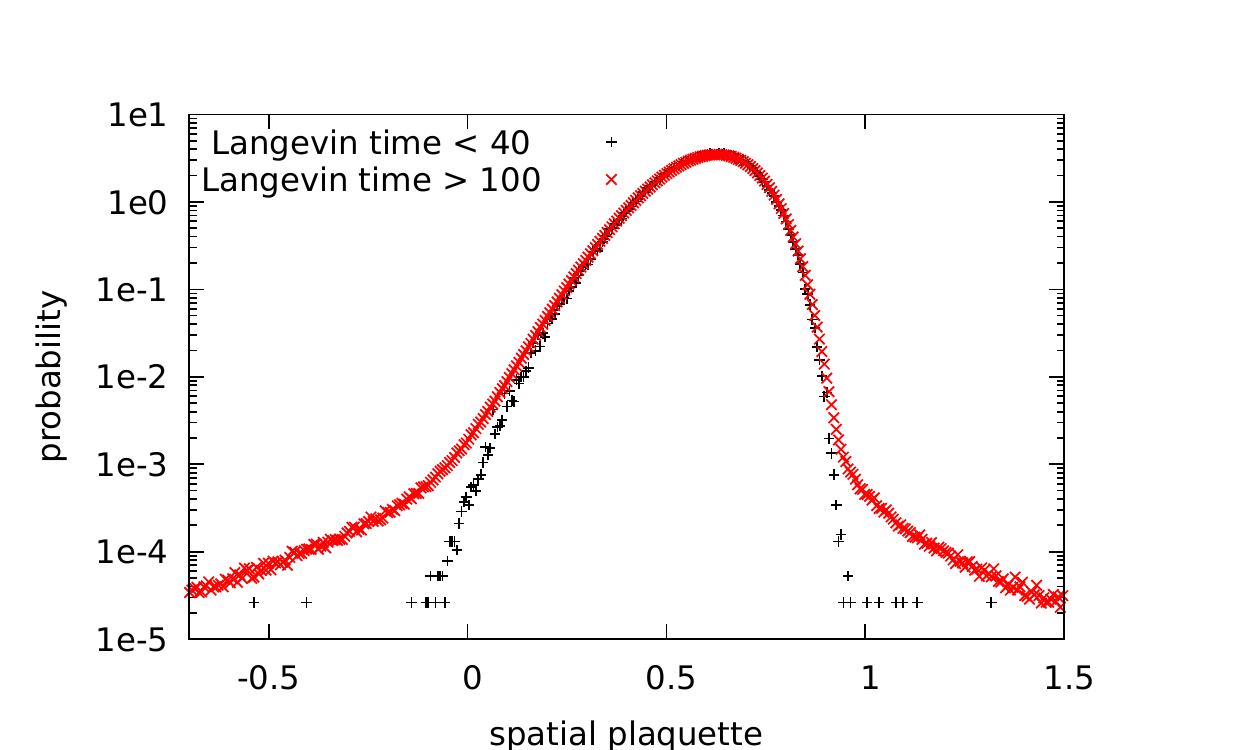}
\includegraphics[width=0.48\textwidth]{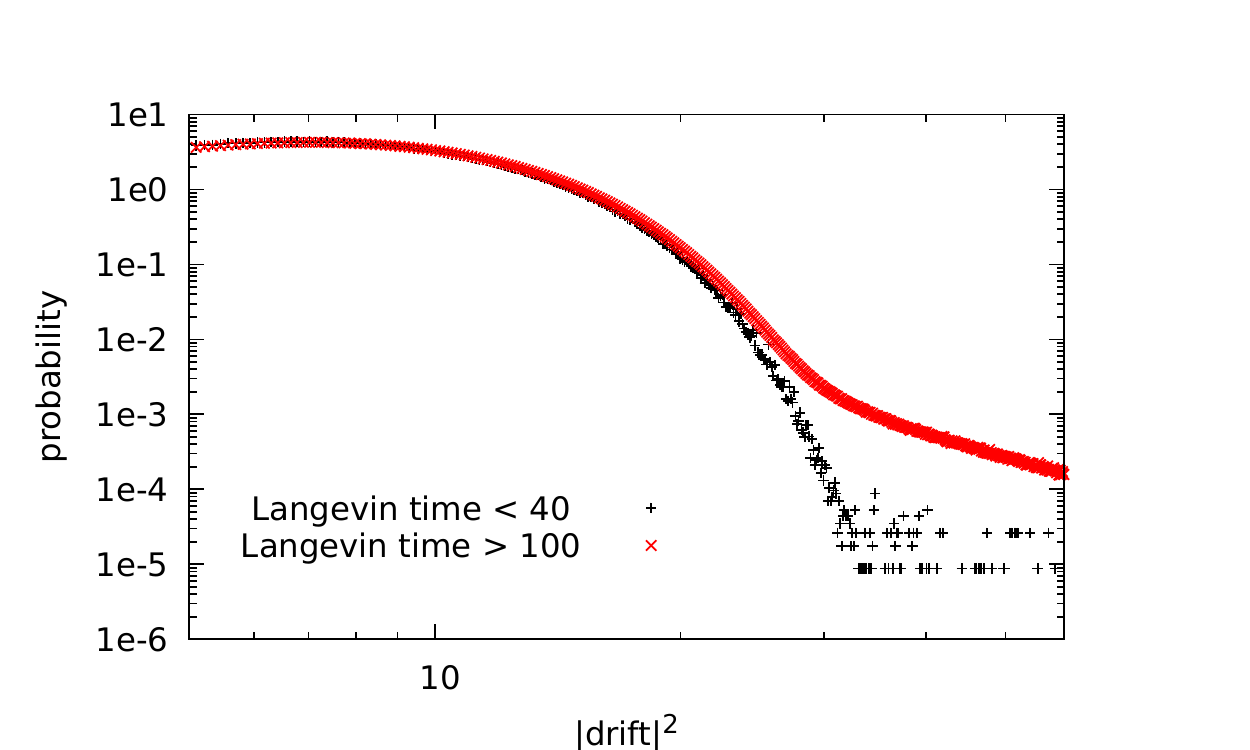}
\caption{ Results in HDQCD at $\beta=6.0, \kappa=0.12, N_F=1, \mu=0.85$ on a $6^4$ lattice.
  Top: Behavior of the spatial plaquette and the unitarity norm as a function of Langevin time.
  Center: Histogram of the spatial plaquette at different time intervals.
  Bottom: Histogram of the absolute value of the drift at different time intervals. }
\label{fig:splaq}
\end{figure}

We investigated the occurrence of boundary terms in HDQCD in \cite{boundaryterms2} and found that as $\beta$ increases at fixed $\kappa$ and $\mu$ the boundary terms tend to become smaller and smaller, thus moving CL so close to the correct value  as obtained from reweighting 
that it becomes indistinguishable from the correct result 
within errorbars. However, even at $\beta=6.0$ if one waits long enough CL becomes slightly wrong.
% in HDQCD.
We show what happens in HDQCD for $V=6^4$, $\beta=6.0$, $\kappa=0.12$, $N_f=1$, $\mu=0.85$ in Fig.~\ref{fig:splaq}. The top plot shows that initially, the observable fluctuates around a plateau at the correct value, which is computed by reweighting. After $t\approx 40$ the fluctuations start to become larger and another equilibrium is reached. This coincides with the average unitarity norm reaching a value of $UN \approx 0.2 $ at $t=40$. Hence, if we cut off the unitarity norm at this value, the observable yields a result consistent with the reweighting value. Note that for this plot some blocking was done, so the points shown are averaged over a small time window.
The idea of using only short times is also supported when looking at the distribution at the observable itself or the criterion from \cite{Nagata:2016vkn}, as seen in the center and bottom plot of figure \ref{fig:splaq}. Here one can see that for long times, the distributions clearly develop tails indicating a failure of CL. However, when only taking short times, where the observable thermalized to the intermediate plateau, there are no tails or only very small tails indicating that CL gives the correct result here.

To summarize, we have established the connection between the UN, the observable's plateau
and the boundary terms \cite{boundaryterms1,boundaryterms2}:
In some cases it is possible to introduce a regularizing term in the action
such that the diffusion toward large UN is suppressed. This leads to
the system staying in the ``plateau'' region asymptotically, such that the boundary
terms vanish, the UN stays small, and the expectation value of the observables
is at the correct value.
 This shows that the correct plateau
of the observable gets its contributions from configurations with small UN, the boundary terms (which spoil correctness) get their contribution from configurations with
large UN, which is reached at large Langevin times, after the system has left the plateau
region. This argument suggests that the existence of an early plateau at small UN provides
the correct expectation value of observables.

The second issue concerns meromorphy of the drift.
The fermionic determinant has zeroes on the complexified
SL(3,$\mathcal{C}$) manifold, which in turn lead to
singularities in the drift terms.
These singularities can in certain cases lead to the
breakdown of the Complex Langevin method. In \cite{Aarts:2017vrv}
it was shown that a sufficient condition for the correctness
of the results is that the zeroes
are outside the distribution 
on the complex manifold.

\begin{figure}
\includegraphics[width=0.48\textwidth]{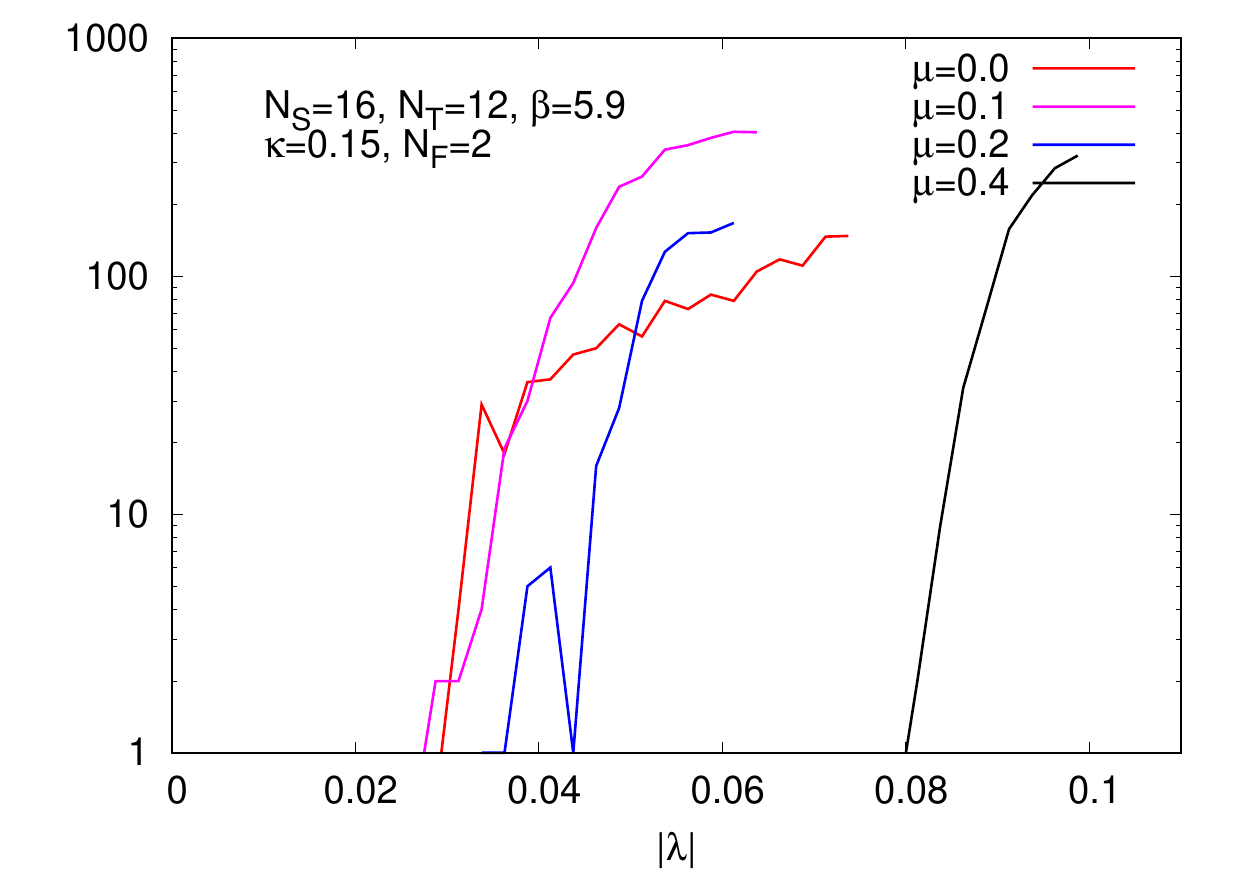}
\caption{The histogram of the absolute value of the smallest eigenvalues
  of the Dirac matrix
  calculated on a  $ 16^3\times 12$ lattice at $ \beta=5.9, \kappa=0.15, N_F=2$.
}
\label{fig:evspect}
\end{figure}

We investigate the eigenvalue distribution of the Dirac matrix to gain an
insight into this question. 
Calculating the whole spectrum of the Dirac matrix is very costly,
and we are interested only in the small eigenvalues which can potentially
cause problems, therefore we use the Krylov-Schur algorithm
to calculate the smallest eigenvalues of the Dirac matrix.
 We find in the interesting region
 close to the transition temperature that for our rather
large masses used in the present investigation 
the spectrum of the Dirac matrix seems to show a very fast
decay at small eigenvalues.
In Fig.~\ref{fig:evspect} we show typical histograms of the
eigenvalues with smallest
absolute values, for various $\mu$ values, on a lattice which is
close to the transition temperature.

\subsection{Towards the continuum limit}
\begin{figure}
\includegraphics[width=0.48\textwidth]{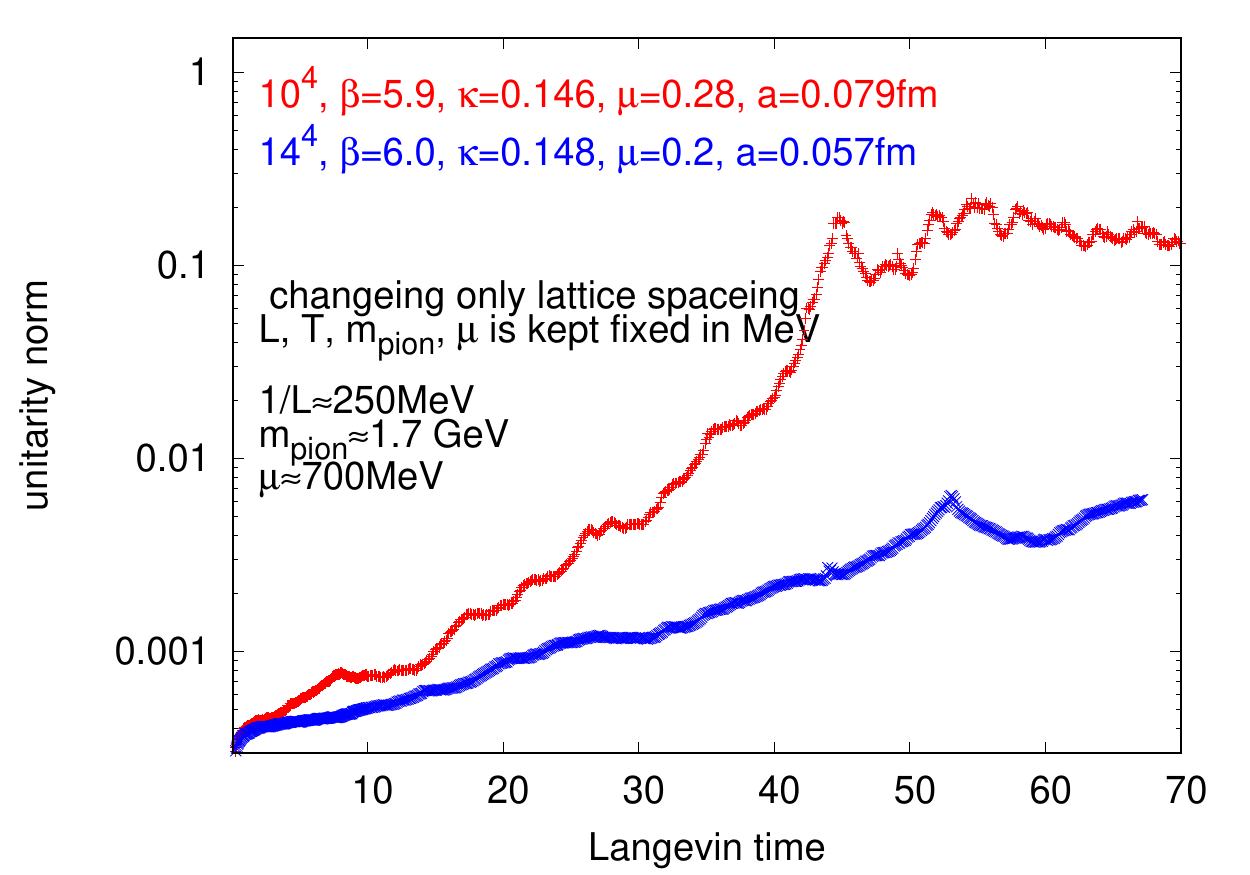}
\includegraphics[width=0.48\textwidth]{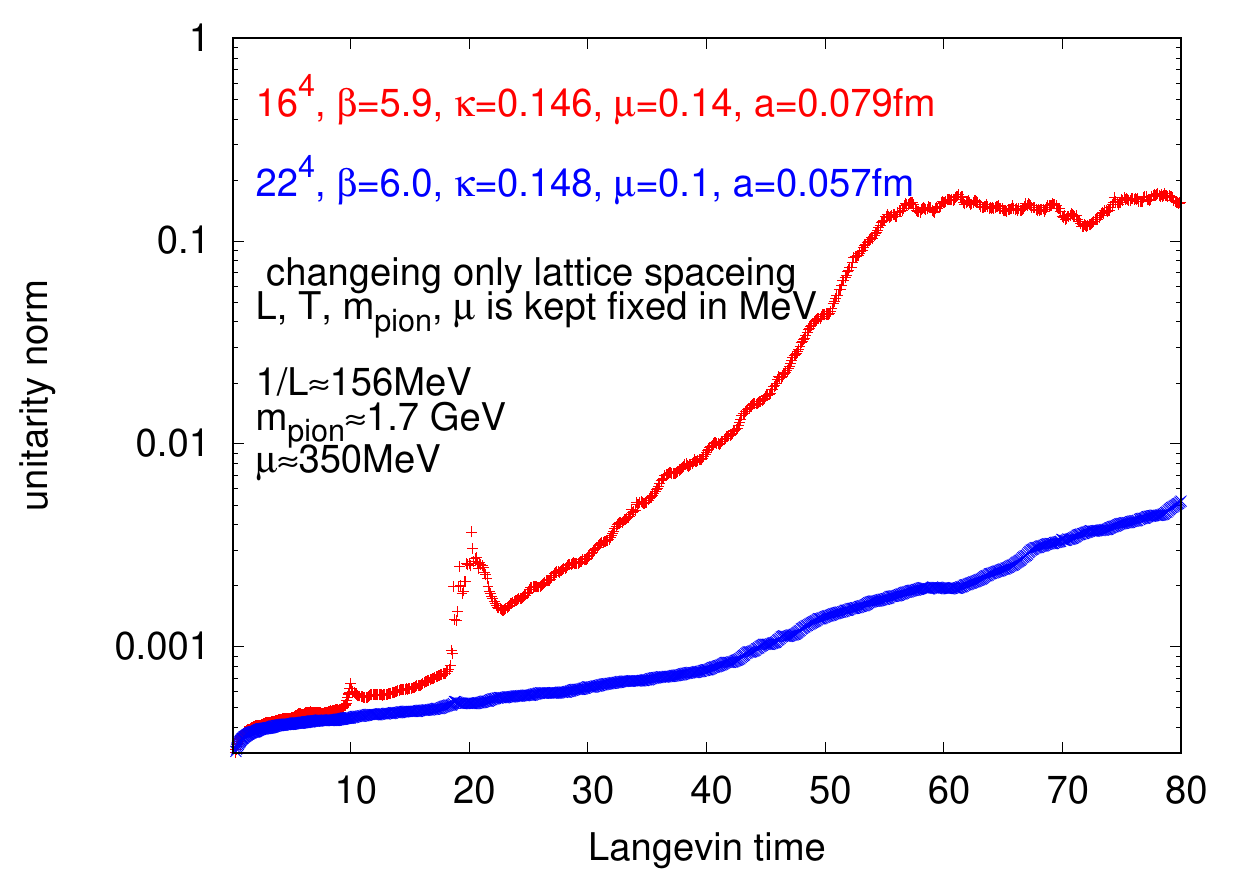}
\caption{Testing the behavior of the Unitarity norm as the continuum limit is approached, using complex Langevin simulations with naive gauge action and Wilson fermions using $N_F=2$ with parameters as indicated.
}
\label{fig:contlim}
\end{figure}

\begin{figure}
\includegraphics[width=0.48\textwidth]{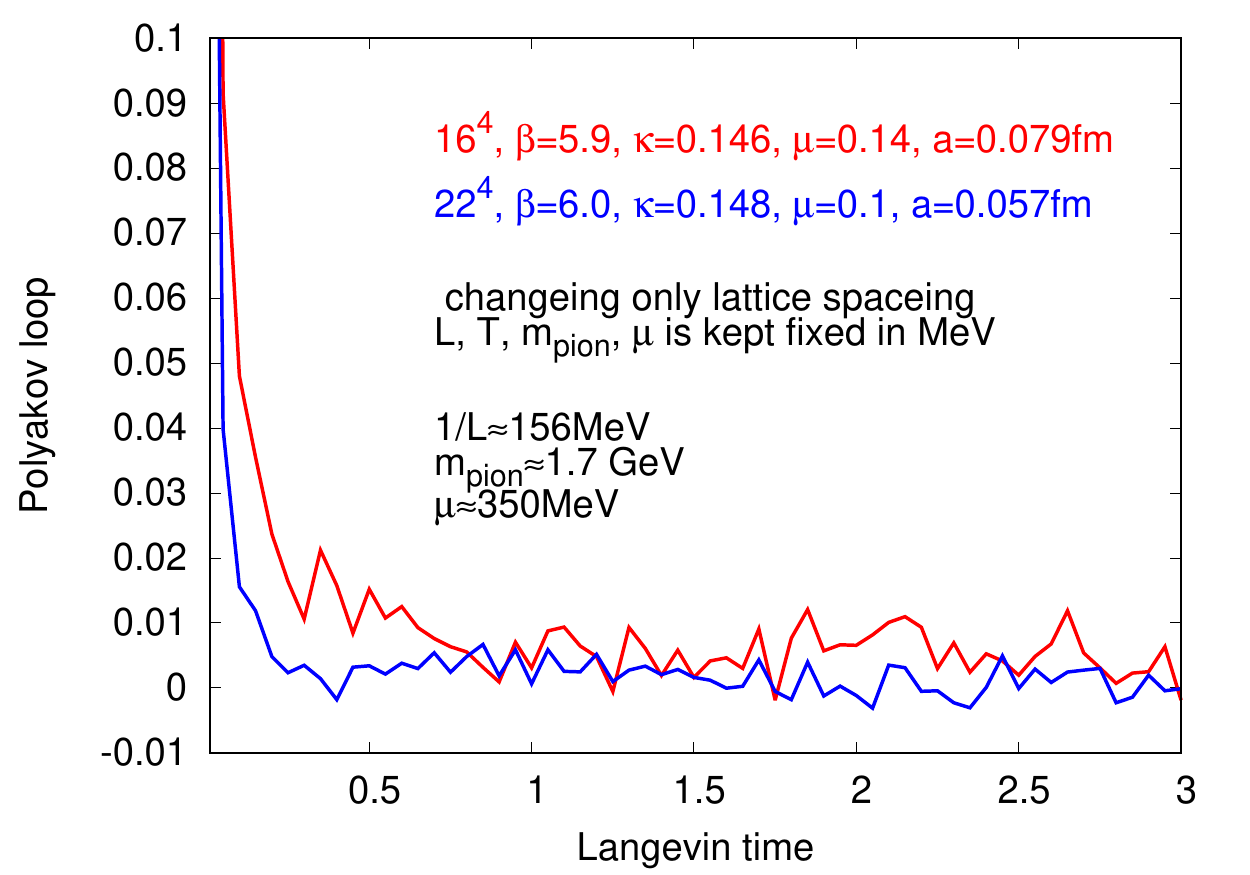}
\caption{The initial thermalization of the Polyakov loop as the continuum limit is approached, using complex Langevin simulations with naive gauge action and Wilson fermions using $N_F=2$ with parameters as in Fig.~\ref{fig:contlim}.}
\label{fig:contlim-therm}
\end{figure}

In this subsection we describe the properties of Complex Langevin simulations
of full QCD as the continuum limit is approached.
One observes that the behavior of the unitarity norm improves closer to the continuum limit.
In Fig.~\ref{fig:contlim} we show typical simulations where the lattice spacing is decreased, while keeping every other quantity fixed in physical units.
The initial thermalization rate as measured using e.g. \nolinebreak
plaquettes or Polyakov loops remains approximately constant, see for the Polyakov loop in Fig.~\ref{fig:contlim-therm}. The autocorrelation times scale approximately with the
lattice spacing, and the rise of the unitarity norm is slower
for the smaller lattice spacing, such that closer to the continuum
there are more statistically
independent samples generated before the unitarity norm grows too high. 

Our strategy thus relies on the system having a short thermalization time
for the physical observables, and a longer thermalization time
for the complexified process, providing a plateau region where
physical observables of interest can be sampled. This strategy
potentially breaks down if the plateau region is not reached before
the fluctuations in the imaginary directions grow large, e.g. around a second
order phase transition or on large lattices.
Closer to the continuum limit however the situation improves: the time window for the plateau region increases.
 Moreover, in the state where the process
has thermalized and there is a discrepancy between complex Langevin
and correct results (caused by boundary terms), 
the discrepancy quickly diminishes as one decreases the lattice
spacing \cite{boundaryterms2}.

Our goal is to scan the transition region of QCD up to large $\mu/T$.
We have seen that for HDQCD the complex Langevin method 
 does not converge to the correct results 
 for small lattice coupling $\beta$.
 We expect a similar behavior for full QCD
 (see \cite{Fodor:2015doa} for a similar behavior with staggered quarks).
Therefore we stay at a safe value for $\beta$ and scan the temperature
 by 
 varying the temporal lattice
extent $N_t$. This allows us to start deep in the confining phase and increase 
the temperature reaching the deconfining phase (where CL simulations already produced results concerning the thermodynamics of QCD \cite{Sexty:2019vqx}). 
 This does not keep the aspect ratio of $N_s/N_t$ intact, which does
 have an effect on some observables as we will see.

We use the plaquette gauge action and unimproved Wilson fermions with $N_F=2$.
To convert to physical units we have measured the lattice spacing
and pion masses, see in Table~\ref{masstable}.
To calculate the drift force for the fermions we use a noisy estimator \cite{Sexty:2013ica}. We use an adaptive algorithm \cite{Aarts:2009dg} to control
the Langevin
stepsize $\epsilon$ in the update such that
$ \textrm {max} (  |K_{ax\nu}| \epsilon) < d $ with the control parameter $d$ typically set to $d \sim 0.001-0.004 $.

\begin{table}
\begin{tabular}{|c|c|c|c|c|}
 \hline
 $\kappa$ & $\beta$ &  $a$(fm) & $ m_\pi a $ & $m_\pi$ (GeV) \\
 \hline
 0.14 & 5.9 &  $0.09152 \pm 0.00045$ & $1.01 \pm 0.0065$ &  $2.18 \pm 0.025$  \\
 0.15 & 5.9 & $0.0655 \pm 0.001$ & $0.4209 \pm 0.017$ &  $1.27 \pm 0.072$  \\ 
 0.14 & 6 & $0.0736 \pm 0.00024$ & $0.9173 \pm 0.0095$ &  $2.46 \pm 0.033$  \\
 0.15 & 6 & $0.05819 \pm 0.00055$ &  $0.2994 \pm 0.013$  & $1.02 \pm 0.054$  \\
\hline 
\end{tabular}
\caption{Pion masses and lattice spacing (defined using the $w_0$ scale as in \cite{Borsanyi:2012zs}) 
 measured measured on a $24^3 \times 48 $ lattice at various $\kappa$ and $\beta$ values. 
The plaquette action is used with naive Wilson fermions at $N_F=2$.}
\label{masstable}
\end{table}

\begin{figure}
\includegraphics[width=0.48\textwidth]{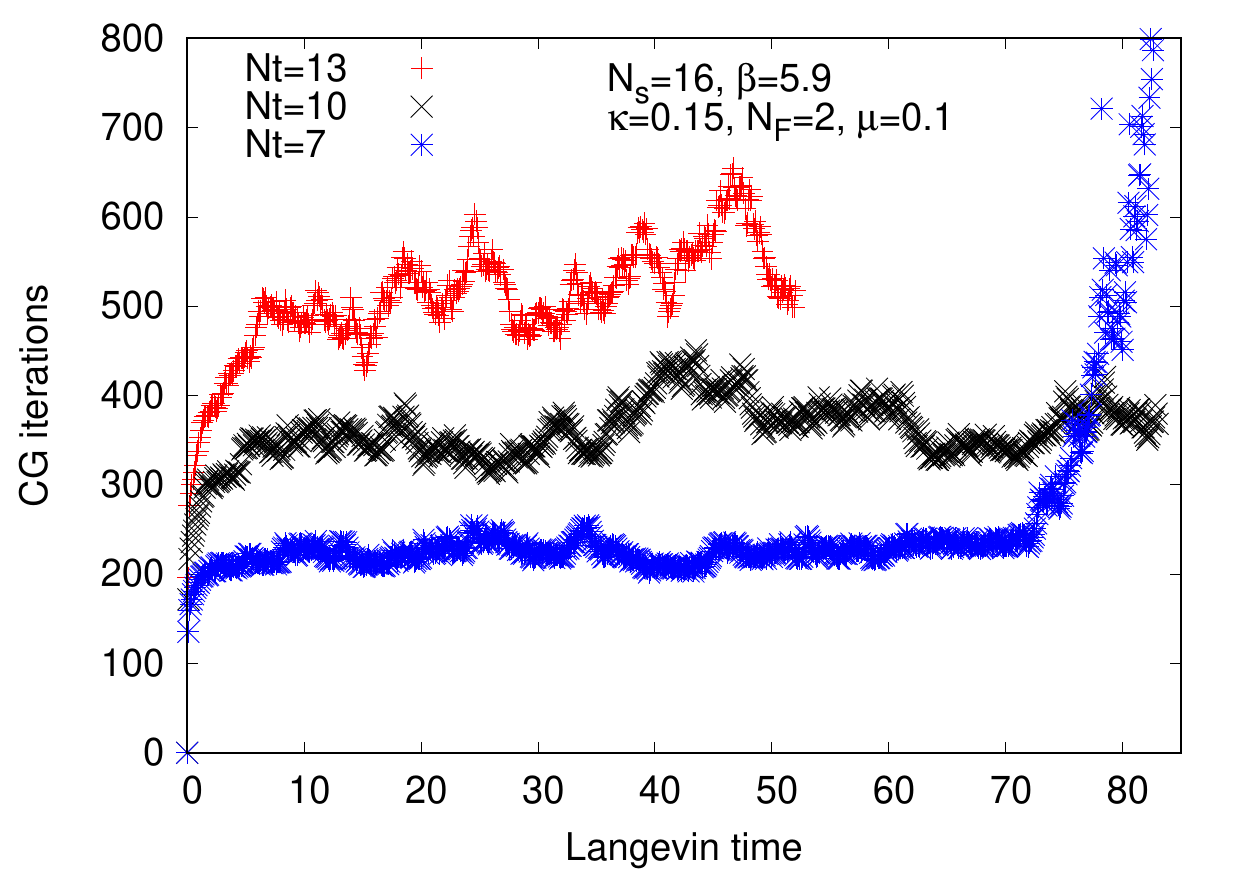}
\includegraphics[width=0.48\textwidth]{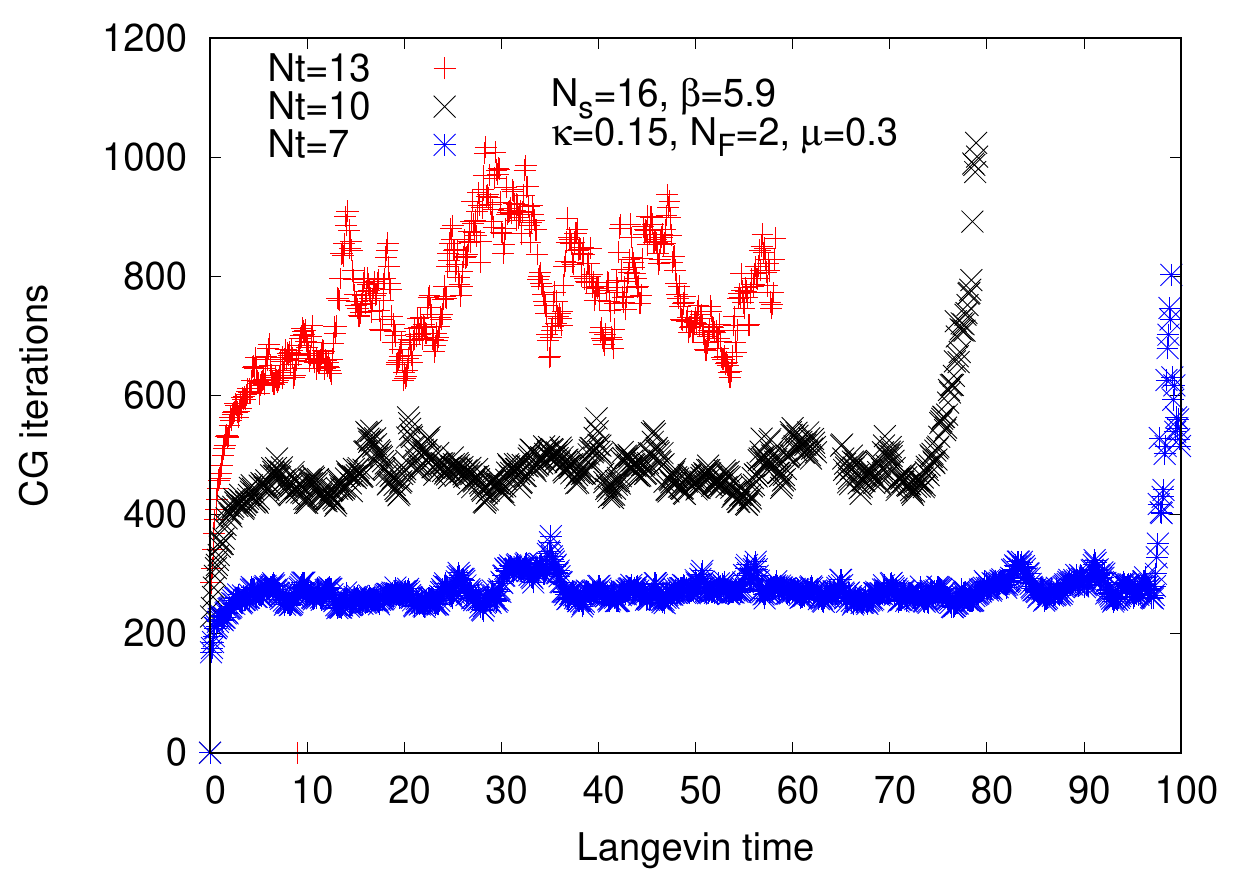}
\caption{The number of required iterations for the calculation of the fermionic drift force for typical runs using the parameters as indicated in the figure. }
\label{fig:CGiter}
\end{figure}

In Fig.~\ref{fig:CGiter} we show the typical number of the required CG iterations used
for the calculation of the fermionic drift terms. The clear upward trend at the end of some of the runs is a consequence of the unitarity norm growing too large, and making the Dirac operator more ill-conditioned. (Some runs were cut short before that happened.) In Hybryd Monte Carlo (HMC) simulations this quantity is normally used to judge the thermalization time of the system, as it is connected to the
lowest eigenmodes of the Dirac operator, which thermalize the slowest. Here some care is required, as the condition number of the Dirac operator is
somewhat sensitive to the unitarity norm, as the non-unitary part of the
link variables typically increases the magnitude of the high eigenvalues and thus increases the condition number. Note also that the spectrum of the Dirac operator might behave in a non-trivial way in CL simulations \cite{Splittorff:2014zca}. Nevertheless the curves suggest that the initial thermalization to a 'meta-stable' state is relatively quick, while the increase of the unitarity norm takes
longer.

\begin{figure}
\includegraphics[width=0.45\textwidth]{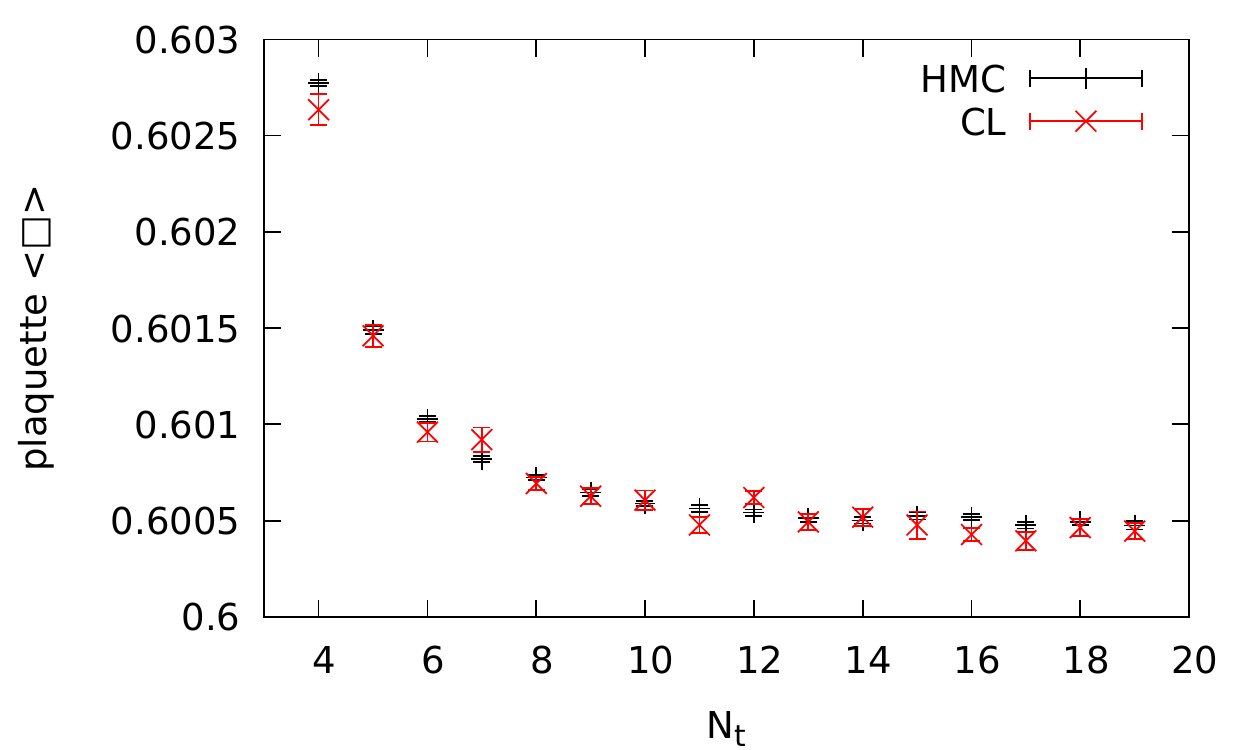}
\includegraphics[width=0.45\textwidth]{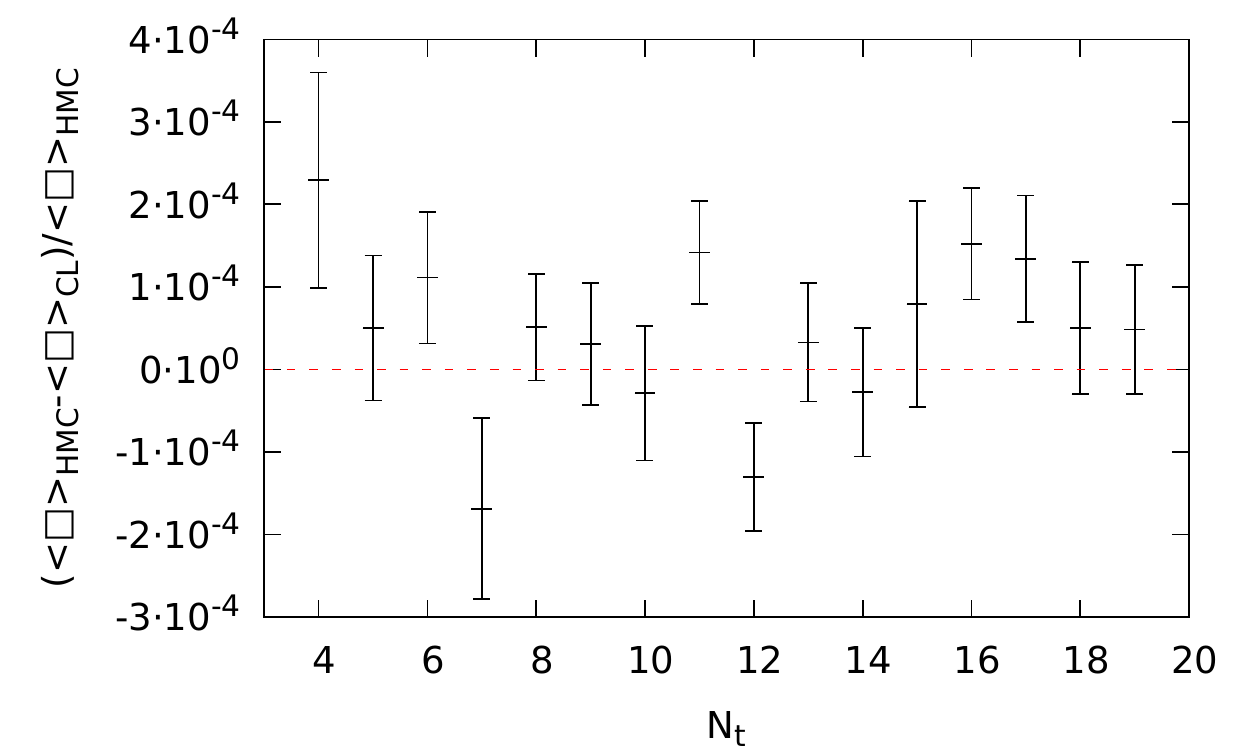}
\caption{Comparison of the plaquette from CL and HMC simulations at $\mu=0$. Top: direct comparison of
 the plaquette, mind the small range of the $y$-axis. Bottom: Relative deviation of the HMC and CL plaquettes.}
\label{fig:compare_HMC_CL}
\end{figure}

It has been noted several times \cite{Bloch:2017jzi,Kogut:2019qmi} that
CL simulations can converge to wrong results even at vanishing
chemical potential $\mu=0$, where the difference to the correct results
quickly decreases with decreasing lattice spacing.
This happens due to the boundary terms at infinity growing large
as the process thermalizes on the complexified manifold
\cite{boundaryterms2}.
However, within our setup we stop the run once the unitarity norm reaches a
 value of $UN\approx 0.1$, in accordance with the plateau region discussed above.
With this procedure we expect to get correct results as long as there
are no physical effects requiring extremely long thermalization such as
a second order phase transition.
In figure \ref{fig:compare_HMC_CL} we compare the plaquette expectation value for a $N_s=12$, $\beta=5.9$, 
$\mu=0$, $\kappa=0.15$ simulation from CL and HMC. One can see that the visible deviations are small and
 agree with zero within statistical errors. The same effect for stout-smeared staggered quarks with $N_f=4$ 
was observed in \cite{Sexty:2019vqx}. Hence, we conclude that within our setup there is practically no
 deviation at $\mu=0$.

\section{Methods and results for the phase transition}
\label{secmethods}

We wish to extract the transition line from different observables.
Natural choices are based on the Polyakov loop, the chiral susceptibility and the density and all their corresponding susceptibilities or higher derivatives.
Since the chiral condensate has proved to be quite noisy, and it
has additive and multiplicative renormalization, here we concentrate
on the Polyakov loop $P$, defined as the spatial average of the trace of temporal loops:
\begin{equation}
 P_\text{bare} = {1\over 3 N_s^3} \sum_{x} \textrm{Tr} \prod_{\tau=0}^{N_\tau-1} U_{x+\tau \hat 0, 0}, 
\end{equation}
with $x$ indexing a spatial slice of the lattice. Similarly we use the average of the trace of the inverse Polyakov loops: 
\begin{equation}
 P'_\text{bare} = {1\over 3 N_s^3} \sum_{x} \textrm{Tr} \left( \prod_{\tau=0}^{N_\tau-1} U_{x,\tau, 0}, \right)^{-1}
\end{equation}
  
The Polyakov loop renormalizes multiplicatively as
 \begin{equation}
 P_\text{ren}=e^{-c(a)N_t}P_ \text{bare}\,, 
 \end{equation}
Hence, if we consider ratios of Polyakov loop observables
renormalization drops out entirely. 
In the following we will not look at the Polyakov loop itself but at a combination of the Polyakov loop and the inverse Polyakov loop
\begin{equation} \label{Pdef}
P=\sqrt{P_\text{bare}P'_\text{bare}}.
\end{equation}
Note that at nonzero chemical potential $ P_\text{bare} \neq (P'_\text{bare})^\dag $, and at finite temperature $P'$ is expected to rise slightly earlier as the chemical potential is increased \cite{gaugecooling}.

The first two of our observables are related to the third order Binder  cumulant
\begin{equation}
B_{3}(\mathcal{O})=\frac{\left<\mathcal{O}^{3}\right>}{\left<\mathcal{O}^{2}\right>^{3/2}}\,.
\end{equation}

We
will look at the following observables built from the Polyakov loop.
\begin{enumerate}
\item $B_3(P-\left<P\right>)$, which takes into account fluctuations properly.
\item $B_3(P)$, which is the third order cumulant for the unsubtracted Polyakov loop. Note that this does not take into account fluctuations properly, but we will show that 
the resulting phase transition line is similar to the first one.
\end{enumerate}

\subsection{The phase transition from $B_3(P-\left<P\right>)$}
\label{sec:sub_cum_3}
\begin{figure}
\centering
\includegraphics[width=0.45\textwidth]{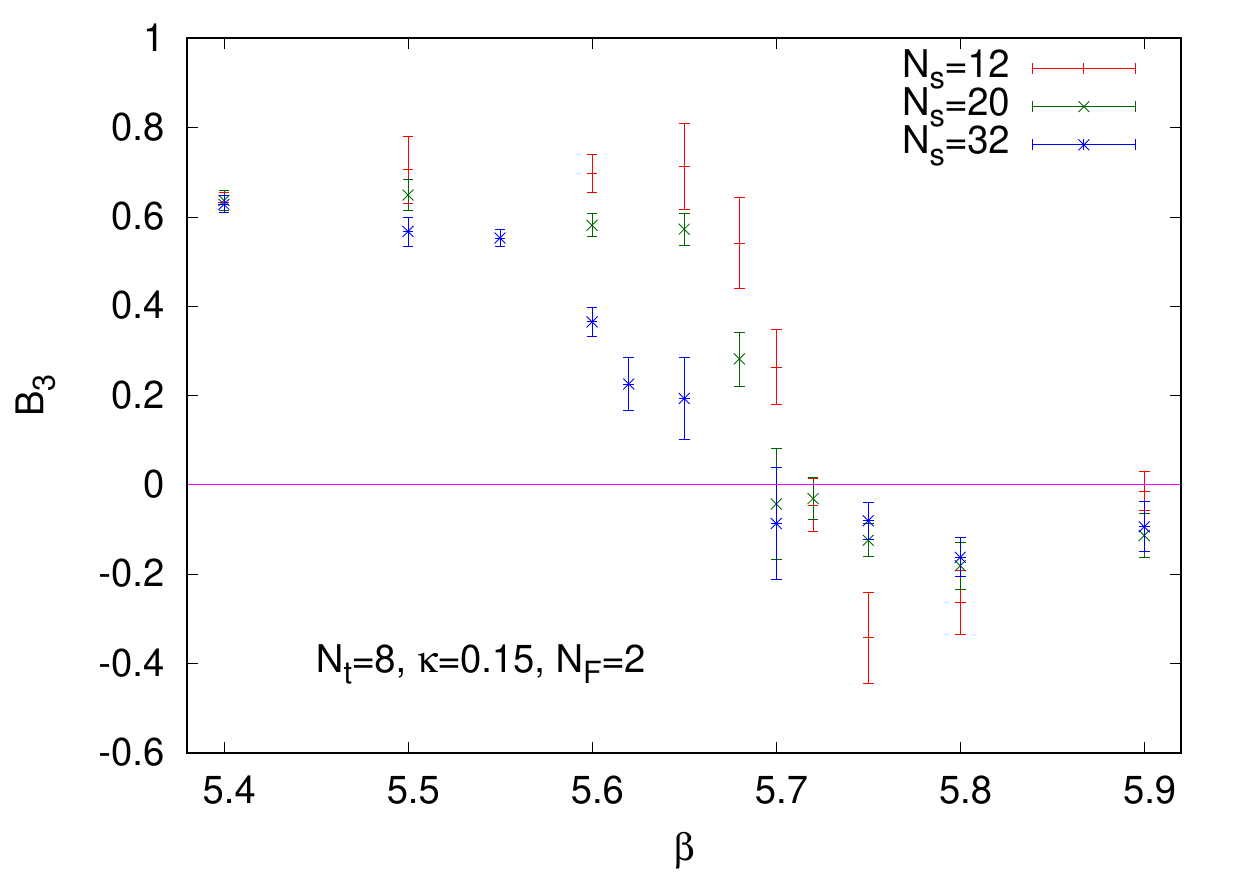}
\includegraphics[width=0.45\textwidth]{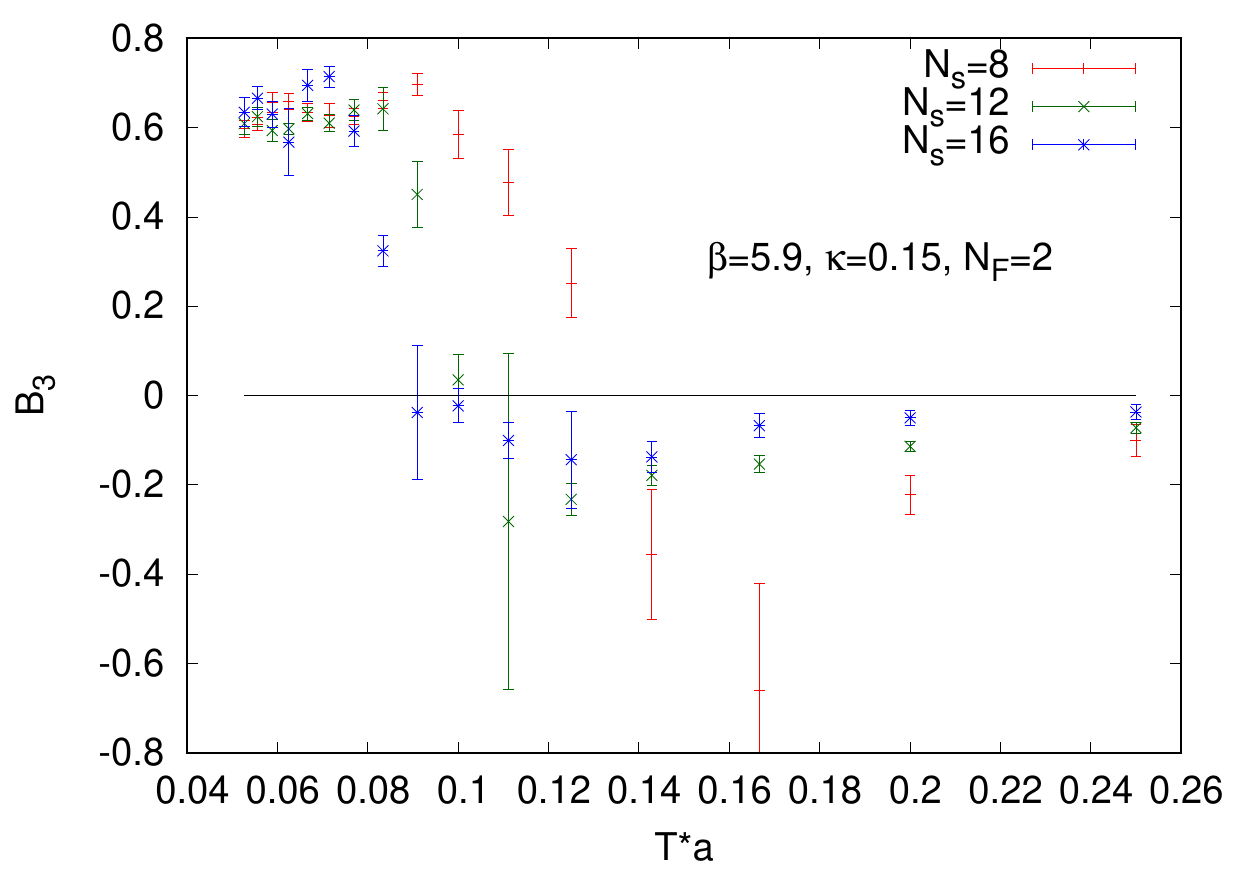}
\caption{Third order cumulant for different volumes as a function of $\beta$ at $N_t=8$ (top) and as a function of $1/N_t=Ta$ at $\beta=5.9$ (bottom).} 
\label{fig:mu0_cum3}
\end{figure}

A standard way to investigate phase transitions is a volume scaling
analysis of cumulants like $B_3(P-\left<P\right>)$ \cite{binder1981finite,Chen:2010ej}.
This analysis works well in the case of an actual phase transition. In QCD we have a crossover instead, so a priori this analysis does not work. However, there are many possibilities to define the phase transition temperature in the case of a crossover transition, see e.g.
\cite{Bazavov:2018mes}. In order to find a criterion, we first look at $\mu=0$ for $\kappa=0.15$, in different volumes. This is visualized in Fig.~\ref{fig:mu0_cum3}. The top plot shows that the usual volume scaling seems to work. The cumulants for different volumes cross at the point where $B_3(P-\left<P\right>)=0$. We use this do define $T_c$ via 
\begin{equation}
B_3(P-\left<P\right>)|_{T=T_c}=0\,.
\end{equation}
For the qualitative understanding of the behavior of $B_3(P-\left<P\right>)$ as a function of $T$ note that $B_3(P-\left<P\right>)$ 
essentially measures the asymmetry of the distribution
of $P$. At large temperatures when $\left<P\right>$ is large a symmetric distribution is
expected. In contrast, at low temperatures $\left< P \right>$ should be small, while the observable $P$ (defined in (\ref{Pdef})) is positive
definite, therefore the distribution of $P$ should be asymmetric.

Note that the bottom plot in figure \ref{fig:mu0_cum3} suggests that this scaling analysis seems not to work when looking at the cumulant as a function of $1/N_t$, however the crossing points get closer as the volume increases. Thus, the criterion can still be used and should be valid in the thermodynamic limit.

We use this method to extract the phase transition temperature at different $\mu$. In practice we apply a linear fit to $B_3(P-\langle P \rangle)$ as a function of $1/N_T$ close to the zero crossing and define $T_c$ from the crossing of the fit function. Statistical errors are calculated using the bootstrap method.
Results are shown in section \ref{sec:results}.
\subsection{The phase transition from $B_3(P)$}
\label{sec:shiftmethod}
One disadvantage of fluctuation quantities such as $B_3(P-\left<P\right>)$ is that they are quite noisy and require high statistics runs. While the method presented in the previous section (\ref{sec:sub_cum_3}) is close to the standard treatment of phase transitions, there are other ways to do that. 
Here, we are interested in the method used in \cite{Endrodi:2011gv,Bonati:2018nut}.
The idea here is the following:
We have an observable $O(T,\mu)$, which is not constant around
the critical temperature (typically it has an inflection point).
We note that $O(T,\mu>0)$ is well
approximated (for small $\mu$ values and close to $T_c(0)$) with $O(T-T_\textrm{shift},\mu=0)$. We then identify $T_\textrm{shift}$ as the shift of the
critical temperature. We formalize this using the following definition:
Provided we know $T_c(\mu=0)$ (using an independent definition e.g. from a peak of a susceptibility),
 we define 
\begin{equation}
\mathcal{O}_c(\mu=0)=\mathcal{O}(\mu=0,T_c(\mu=0))\,,
\end{equation}
 and define the phase transition temperature via
 \begin{equation}
\mathcal{O}(T_c,\mu)= \mathcal{O}_c(\mu=0)\,.
 \end{equation}
Note that this procedure relies very much on a precise determination of $T_c(\mu=0)$, which can be performed using HMC simulations, which are typically faster than CL simulations and thus allow for more statistics.

To define $T_c$ in practice we have used spline interpolation to define a continous function for  $B_3(P)$ as a function of $1/N_t$. Similarly to
the first definition, statistical errors are calculated
using the bootstrap method.
Again, we will show results of this procedure in section \ref{sec:results}.

\subsection{Results}

\label{sec:results}
In this section, we show numerical results of our simulations.
We have used the plaquette gauge action with two flavours of naive Wilson
fermions, at $\beta=5.9$ and $\kappa=0.15$. This corresponds
to a relatively heavy pion mass of $\approx 1.3 $ GeV, as seen in
Table~\ref{masstable}.
We use the temporal extent
of the lattice to vary the temperature, and we show results for two different
spatial volumes $N_s=12,16$, as smaller volumes proved to be too noisy.

\subsubsection{The cumulants}
\begin{figure}[h]
\includegraphics[width=0.45\textwidth]{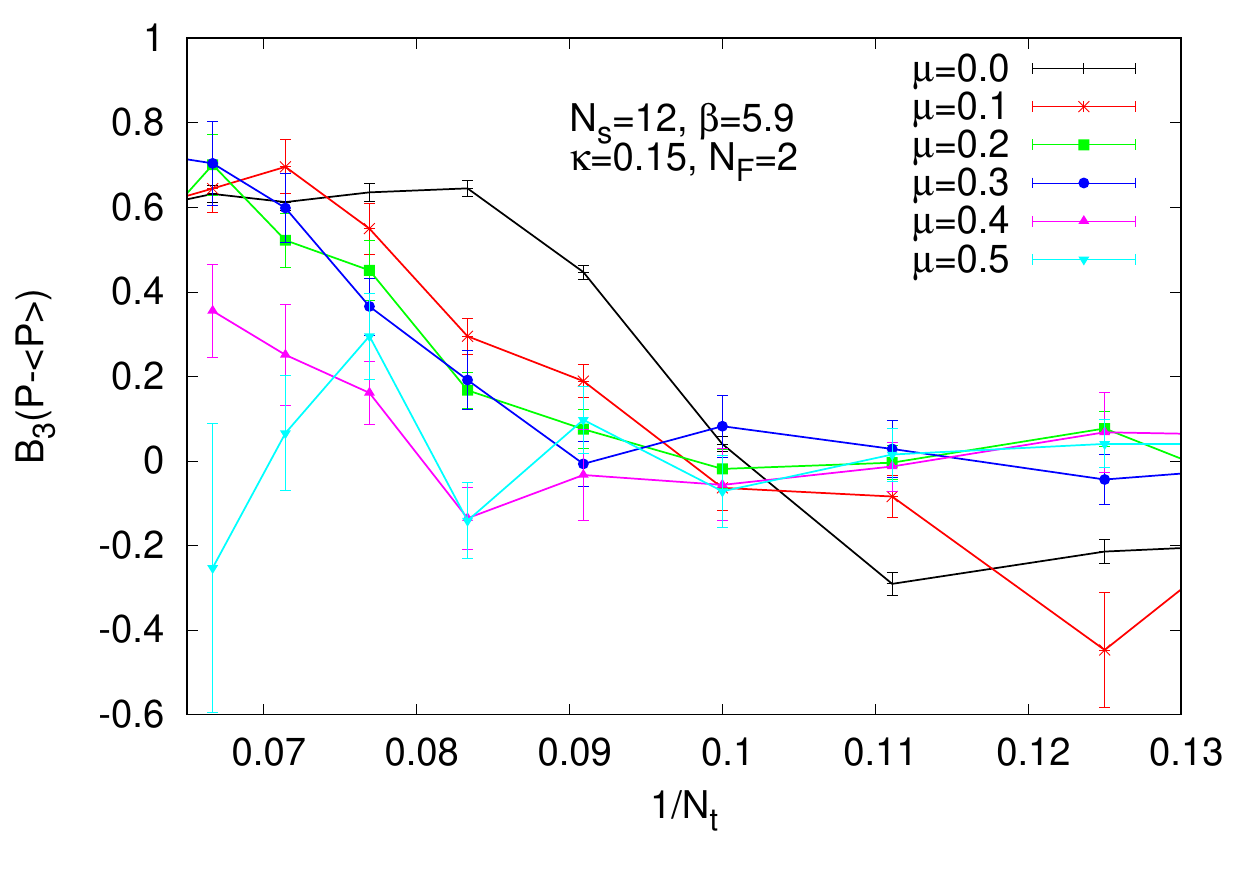}
\includegraphics[width=0.45\textwidth]{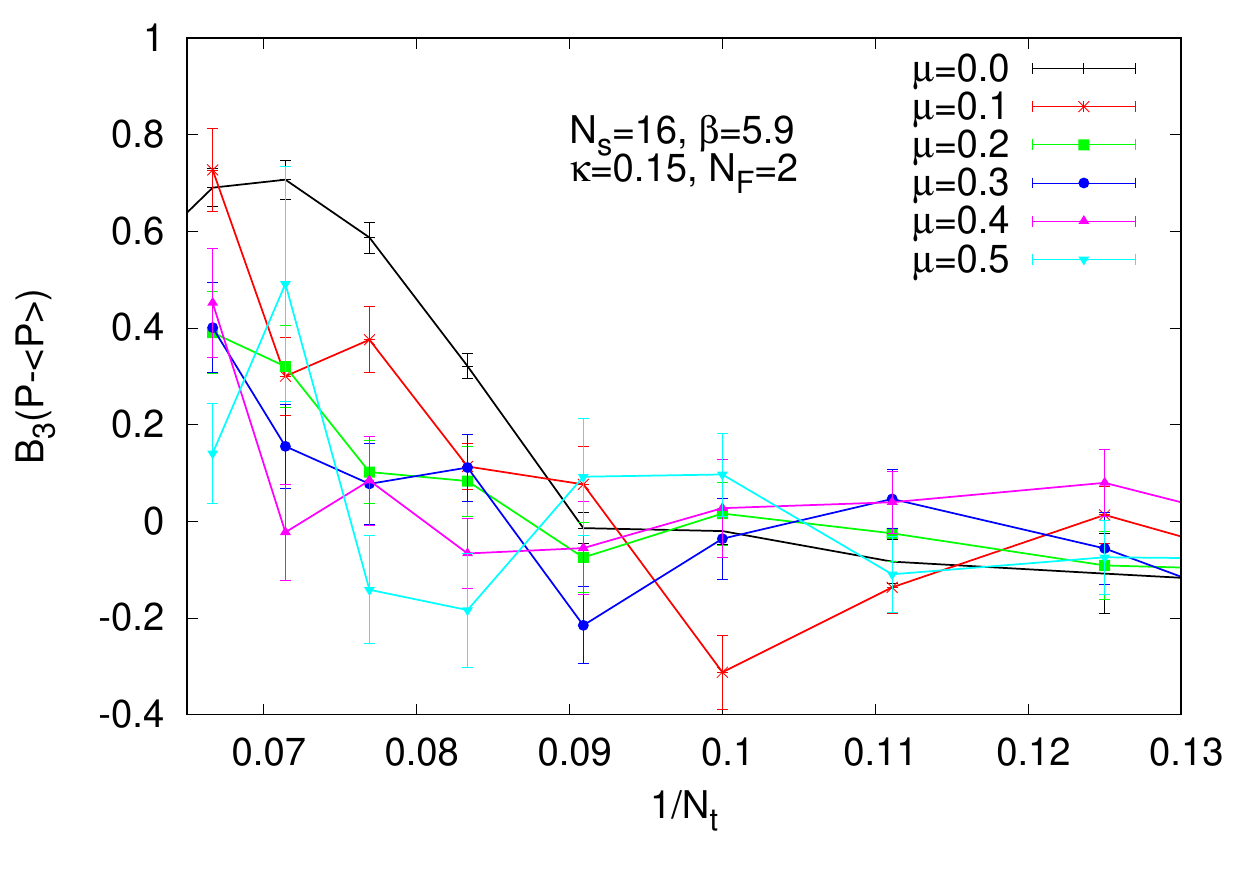}
\caption{$B_3(P-\left<P\right>)$ as a function of $1/N_t$ for $N_s=12$ (top) and $N_s=16$ (bottom). Only the region of the zero crossing is shown.}
\label{fig:cum3_sub}
\end{figure}

\begin{figure}[h]
\includegraphics[width=0.45\textwidth]{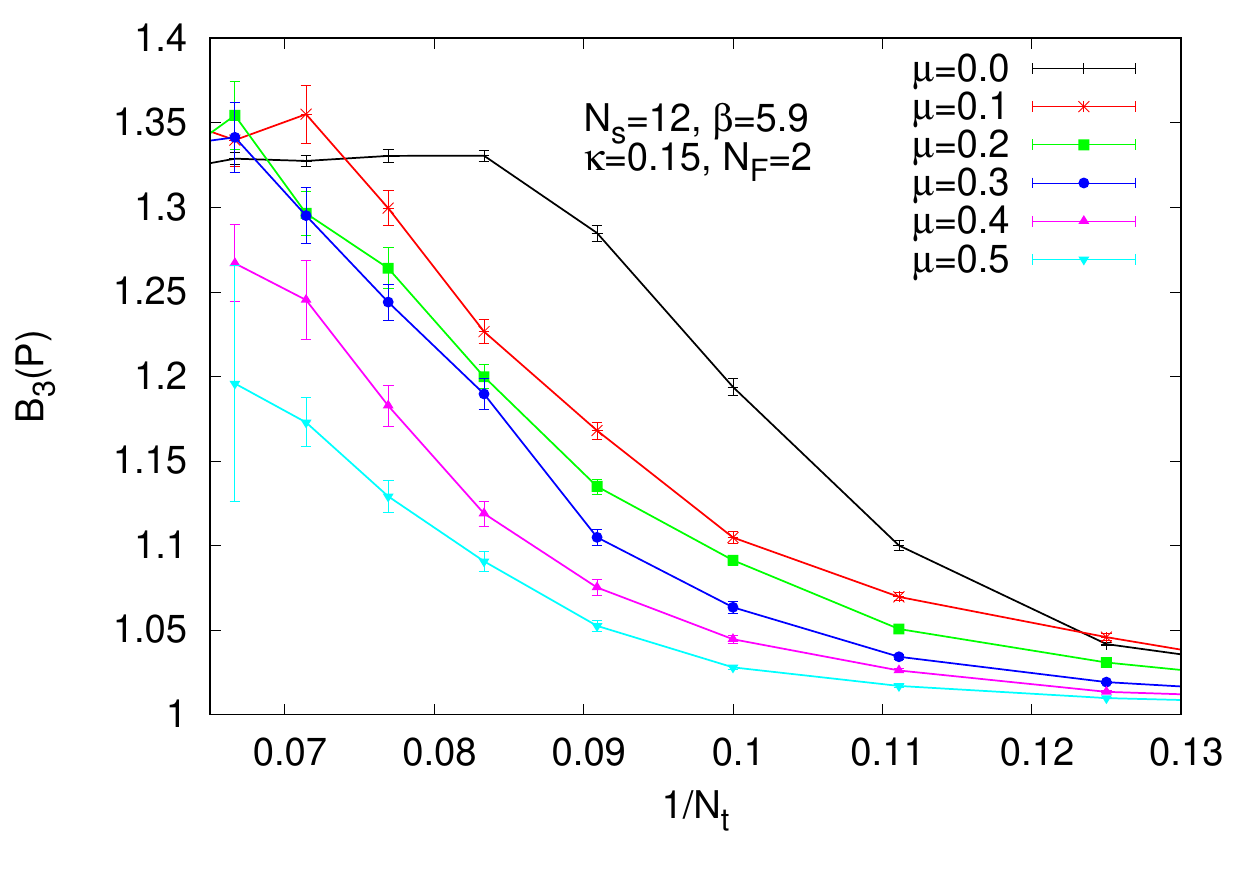}
\includegraphics[width=0.45\textwidth]{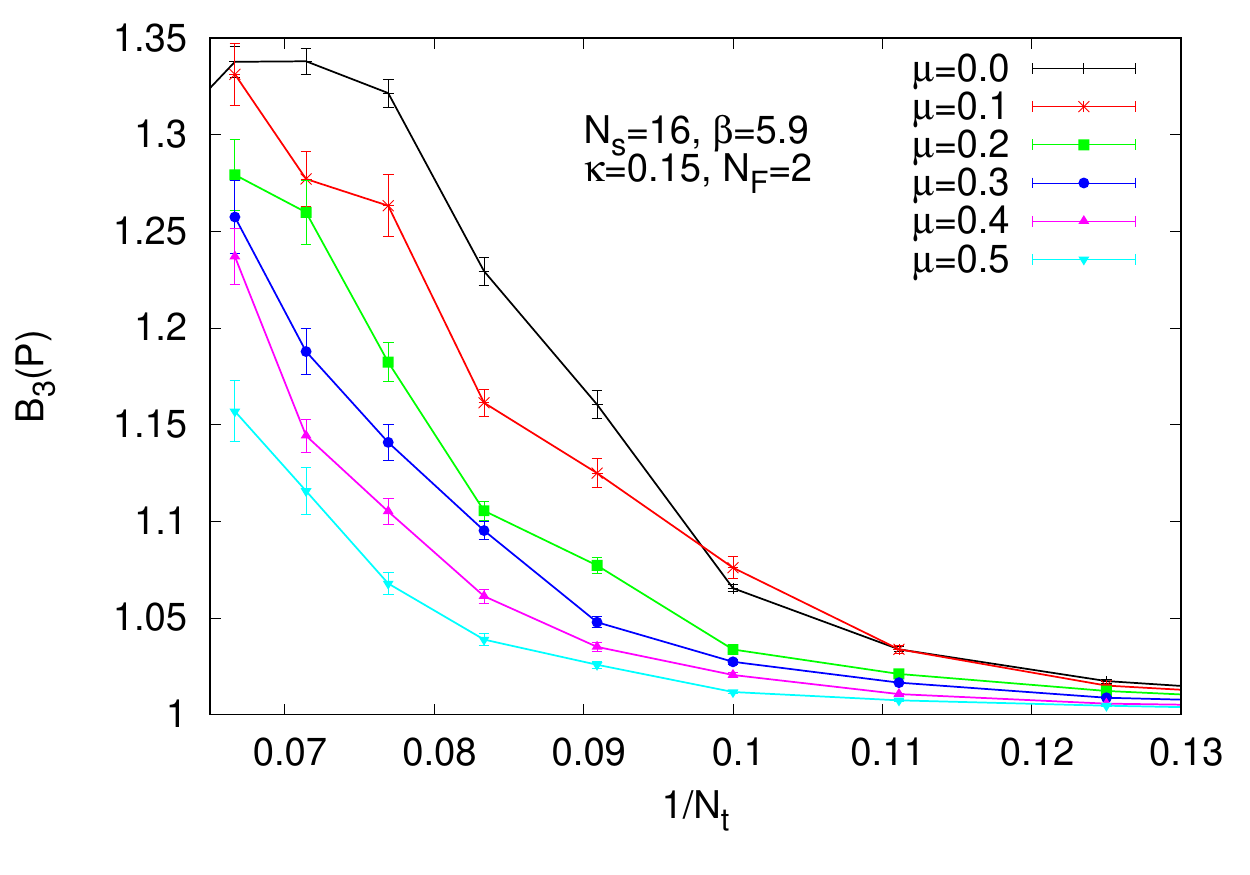}  
\caption{$B_3(P)$ as a function of $1/N_t$ for $N_s=12$ (top) and $N_s=16$ (bottom). Only the region of interest is shown.}
\label{fig:cum3_unsub}
\end{figure}

We show $B_3(P-\left<P\right>)$  for $N_s=12,16$ in Fig.~\ref{fig:cum3_sub} and $B_3(P)$  for the same volumes in Fig.~\ref{fig:cum3_unsub}.


We find that the simulations are much noisier for smaller $\mu$, while the curves become much smoother and better behaved for larger $\mu$.
We find that in the remaining analysis it is hard to extract a transition temperature at $\mu=0$ due to too low statistics. Instead, in the following we use $T_c$ from the HMC analysis, see Fig.~\ref{fig:mu0_cum3}. Note however, that there is no discrepancy between HMC and CL, as can be seen in Fig.~\ref{fig:compare_HMC_CL}.\\
\subsubsection{Phase transition temperatures and curvature of the transition line}
Once we have extracted the transition temperatures in all cases, we can compute the curvature of the transition line. This is a standard value regularly computed using lattice simulations, however in conventional lattice simulations it is only accessible around $\mu=0$ using Taylor expansion, imaginary chemical potentials or reweighting.
\begin{figure}[h]
\includegraphics[width=0.48\textwidth]{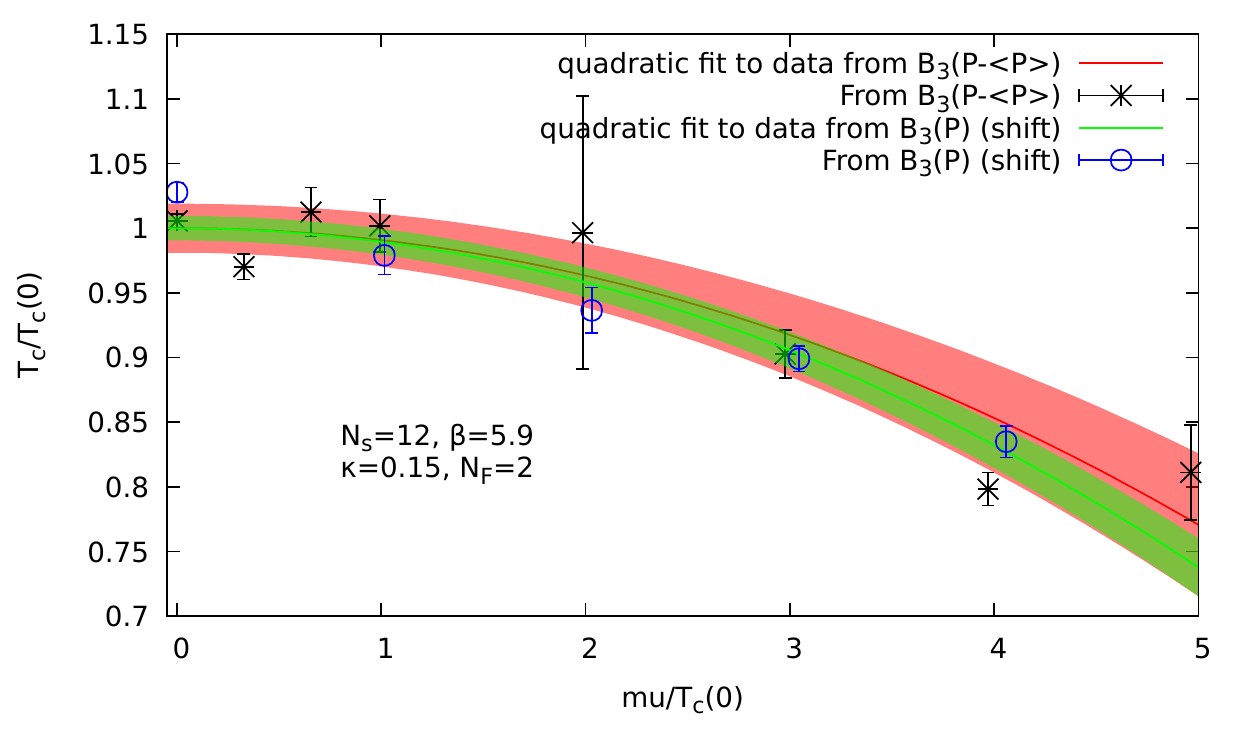}
\includegraphics[width=0.48\textwidth]{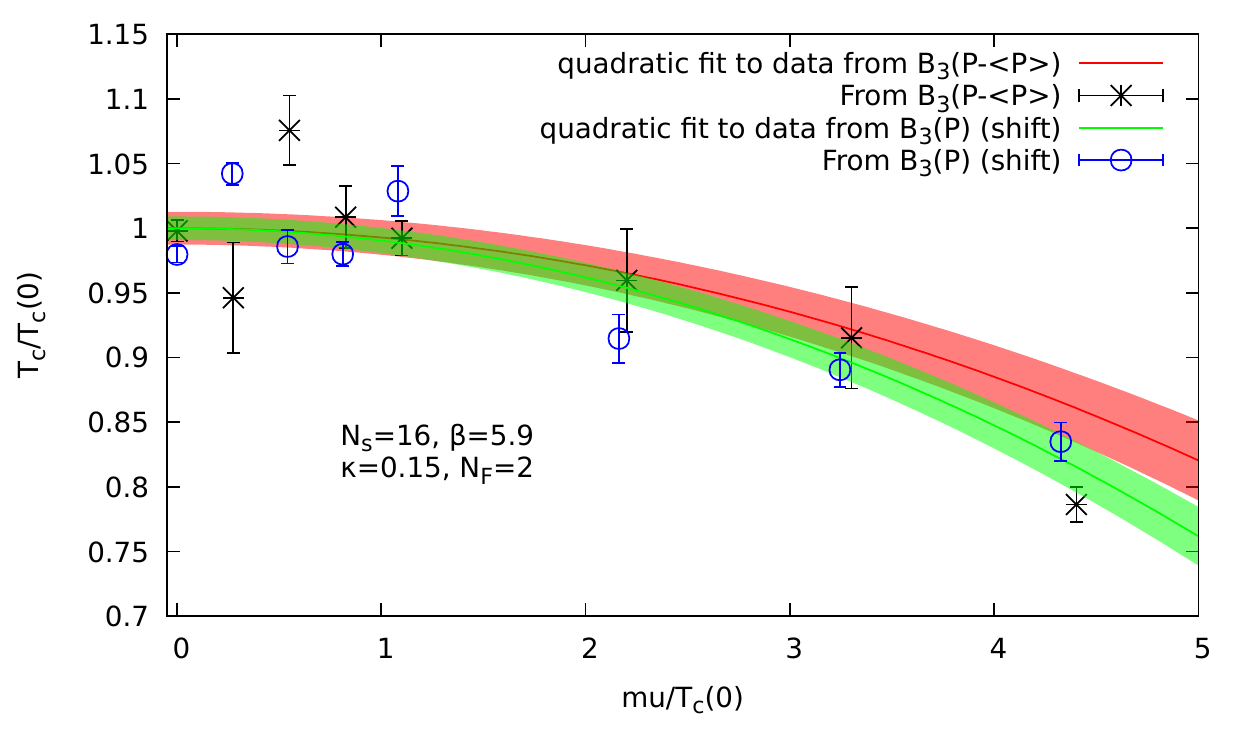}
\caption{Critical temperature as a function of chemical potential from different methods. Top: $N_s=12$; Bottom: $N_s=16$}
\label{fig:Transition_lines}
\end{figure}
We fit our results with a quadratic function of the form
\begin{equation}
T_c(\mu)=T_c(0)-\kappa_2 \frac{9 \mu^2}{T_c(0)}\,,
\label{eq:fitfunc}
\end{equation}
and after the fit normalize to $T_c(\mu)/T_c(0)$. 
We show the resulting phase transitions in Fig.~\ref{fig:Transition_lines} for two different volumes, note that both axes have been normalized with $T_c(0)$, i.e. the parameter that comes from the fit of equation \eqref{eq:fitfunc} to the data.

\subsubsection{Comparison of the methods to define $T_c(\mu)$}
Results of the different methods are compared in table \ref{tab:curvature}.
We used a value of $a=0.0655(1)\text{fm}$ for $\kappa=0.15$
%and $a=0.07913(53)$ for $\kappa=0.126$
measured via gradient flow using the $w_0$ scale to convert the transition temperature into physical units, see also in Table~\ref{masstable}.
One notes a good agreement between the different methods of the definition
of the transition temperature. Finite volume effects are especially visible
in the value of $T_c(0)$.

\begin{figure}
\centering
\includegraphics[width=0.45\textwidth]{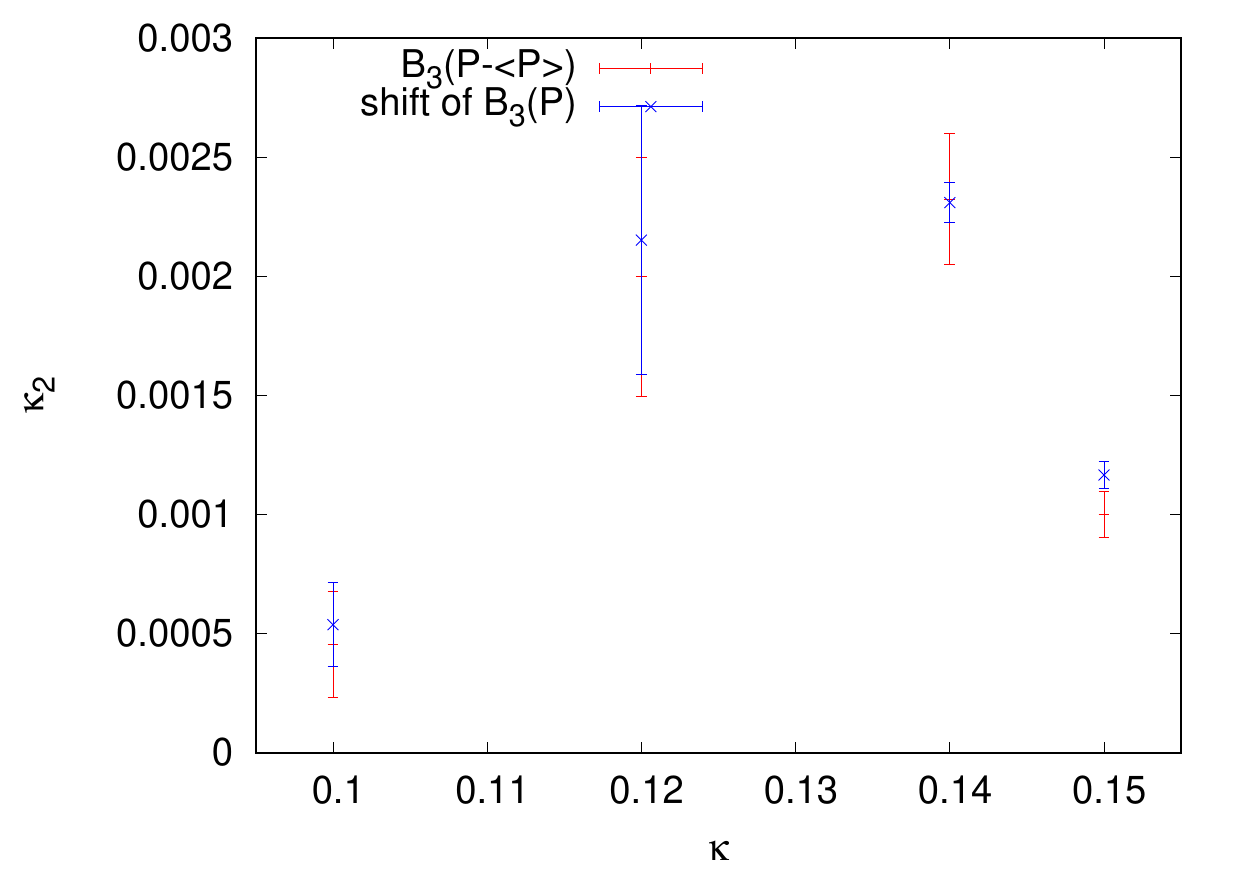}
\caption[Density susceptibility]{The curvature parameter $\kappa_2$ as a
  function of the $\kappa$ parameter of the fermion action as measured
  on lattices with $N_s=12$ and $\beta=5.9$.}
\label{fig:kappavskappa}
\end{figure}

\begin{table}[h]
\begin{tabular}{ |l |c| c| c| c |}
\hline
Method & $N_s$&  $\kappa_2$& $T_c(0)\times a$ & $T_c(0)/\text{MeV}$\\
\hline

fit $B_3(P-\left<P\right>)$ & 12 & $0.001002(96)$ & $0.1004(8)$ & $303(2)$ \\
\hline

shift $B_3(P)$ & 12& $0.001167(55)$ & $0.0987(9)$ & $297(3)$ \\
\hline

fit $B_3(P-\left<P\right>)$ & 16&$8.1(2.4)10^{-4}$ & $0.091(3)$ & $270(10)$ \\
\hline

shift $B_3(P)$ & 16& $0.001042(53)$ & $0.0926(9)$ & $279(3)$ \\
\hline
\end{tabular}
\caption[Results for the curvature of the QCD transition line over the whole range and critical temperature at $\mu=0$]
        {The fitted curvature and $T_c(0)$ according to (\ref{eq:fitfunc})
          where $T_c(\mu)$ is obtained from different methods:
          the zero crossing of  $B_3(P-\left<P\right>)$ as described in
          Sec. \ref{sec:sub_cum_3} and
          the 'shift method' described in Sect. \ref{sec:shiftmethod}.
          The curvature is given for $\mu_B=3\mu$, 
        see Eq. (\ref{mubarionkappa}).
    The column $T_c(0)$ is the parameter resulting from the quadratic fit.}
\label{tab:curvature}
\end{table}
In Fig.~\ref{fig:kappavskappa} we show the curvature $\kappa_2$ as function
of the $\kappa$ parameter in order to ascertain its dependence on the
pion mass. As expected, at very high fermionic mass (corresponding
to a small $\kappa$ parameter) we observe a small $\kappa_2$ parameter,
as also seen in a strong coupling and hopping parameter expansion \cite{Fromm:2011qi},
and in an earlier study of HDQCD \cite{hdqcdpd}.
Curiously, we observe a non-monotonic behavior showing a maximum
at intermediate masses. 

\section{Conclusions}
\label{secconc}

In this paper we have studied the deconfinement transition of QCD
for a range of chemical potentials up to $\mu/T \sim 5$, using the
plaquette gauge action with naive Wilson fermions at $N_F=2$.
To circumvent the sign problem at $\mu>0$ we have used the Complex Langevin
equation.

To stabilize the simulation we use adaptive step size and gauge cooling.
To limit the effects from the boundary terms spoiling the
correctness proof we monitor the unitarity norm UN (\ref{UN}).
Following the results from simple models and from HDQCD we
read the observables' averages from the thermalized plateau
developing at small UN. We also made sure that our simulation does
not go near the zeroes of the determinant.

We have defined the transition temperature using either the zero crossing
of the third order Binder cumulant of the Polyakov loop, as well
as a 'shift method' where we assume that the temperature dependence
of some observable $B(T)$ changes to $B(T-T_\textrm{shift})$ at nonzero $\mu$
(in a range close to $T_c$), which then allows the definition of the
shifted $T_c(\mu)$.

We have carried out most simulations at $\beta=5.9$, $\kappa=0.15$ which correspond to
a relatively heavy $m_\pi \approx 1.3$ GeV, at spatial
volumes $N_s=12$ and $N_s=16$.
The measured transition temperatures are well described with a quadratic
dependence on $\mu$ up to $\mu \sim 5 T $.
We observe a good agreement between various definitions of the
transition temperature.

The determination of the
transition temperature with the Binder cumulant works relatively well, in
spite of the transition being a smooth crossover.
This might signal that a second order transition is nearby.
A candidate for that is the transition line at the upper right
corner of the Columbia plot \cite{Brown:1990ev}, where the deconfinement transition
is expected to turn first order at heavy
quark masses. To investigate the applicability of the method
further studies are needed at smaller quark masses and larger
spatial volumes.

\acknowledgments \noindent

We thank Erhard Seiler for discussions and interest beyond the common work
(see Sec.~\ref{secsetupCL}). We thank Gert Aarts, Philippe De Forcrand, Jan Pawlowski and Felix Ziegler
for discussions. M. Scherzer thanks Jun Nishimura for discussions.
D.~Sexty is funded by the 
Heisenberg programme of the DFG (SE 2466/1-2). M.~Scherzer and I.-O.~Stamatescu are supported by the DFG under grant STA283/16-2. The authors acknowledge support by the High Performance and Cloud Computing
Group at the Zentrum f\"ur Datenverarbeitung of the University of T\"ubingen,
the state of Baden-W\"urttemberg through bwHPC and the German Research
Foundation (DFG) through grant no INST 37/935-1 FUGG.
We also acknowledge the Gauss Centre for Supercomputing (GCS) for
providing computer time on the supercomputers JURECA/BOOSTER and JUWELS
at the J\"ulich Supercomputing Centre (JSC) under the
GCS/NIC project ID HWU32.
Some parts of the numerical calculations were done on the GPU cluster at the University of Wuppertal.

\bibliography{mybib}

\begin{thebibliography}{49}%
\makeatletter
\providecommand \@ifxundefined [1]{%
 \@ifx{#1\undefined}
}%
\providecommand \@ifnum [1]{%
 \ifnum #1\expandafter \@firstoftwo
 \else \expandafter \@secondoftwo
 \fi
}%
\providecommand \@ifx [1]{%
 \ifx #1\expandafter \@firstoftwo
 \else \expandafter \@secondoftwo
 \fi
}%
\providecommand \natexlab [1]{#1}%
\providecommand \enquote  [1]{``#1''}%
\providecommand \bibnamefont  [1]{#1}%
\providecommand \bibfnamefont [1]{#1}%
\providecommand \citenamefont [1]{#1}%
\providecommand \href@noop [0]{\@secondoftwo}%
\providecommand \href [0]{\begingroup \@sanitize@url \@href}%
\providecommand \@href[1]{\@@startlink{#1}\@@href}%
\providecommand \@@href[1]{\endgroup#1\@@endlink}%
\providecommand \@sanitize@url [0]{\catcode `\\12\catcode `\$12\catcode
  `\&12\catcode `\#12\catcode `\^12\catcode `\_12\catcode `\%12\relax}%
\providecommand \@@startlink[1]{}%
\providecommand \@@endlink[0]{}%
\providecommand \url  [0]{\begingroup\@sanitize@url \@url }%
\providecommand \@url [1]{\endgroup\@href {#1}{\urlprefix }}%
\providecommand \urlprefix  [0]{URL }%
\providecommand \Eprint [0]{\href }%
\providecommand \doibase [0]{http://dx.doi.org/}%
\providecommand \selectlanguage [0]{\@gobble}%
\providecommand \bibinfo  [0]{\@secondoftwo}%
\providecommand \bibfield  [0]{\@secondoftwo}%
\providecommand \translation [1]{[#1]}%
\providecommand \BibitemOpen [0]{}%
\providecommand \bibitemStop [0]{}%
\providecommand \bibitemNoStop [0]{.\EOS\space}%
\providecommand \EOS [0]{\spacefactor3000\relax}%
\providecommand \BibitemShut  [1]{\csname bibitem#1\endcsname}%
\let\auto@bib@innerbib\@empty
%</preamble>
\bibitem [{\citenamefont {Petreczky}(2012)}]{Petreczky:2012rq}%
  \BibitemOpen
  \bibfield  {author} {\bibinfo {author} {\bibfnamefont {P.}~\bibnamefont
  {Petreczky}},\ }\href {\doibase 10.1088/0954-3899/39/9/093002} {\bibfield
  {journal} {\bibinfo  {journal} {J. Phys.}\ }\textbf {\bibinfo {volume}
  {G39}},\ \bibinfo {pages} {093002} (\bibinfo {year} {2012})},\ \Eprint
  {http://arxiv.org/abs/1203.5320} {arXiv:1203.5320 [hep-lat]} \BibitemShut
  {NoStop}%
%%CITATION = ARXIV:1203.5320;%%
\bibitem [{\citenamefont {Philipsen}(2013)}]{Philipsen:2012nu}%
  \BibitemOpen
  \bibfield  {author} {\bibinfo {author} {\bibfnamefont {O.}~\bibnamefont
  {Philipsen}},\ }\href {\doibase 10.1016/j.ppnp.2012.09.003} {\bibfield
  {journal} {\bibinfo  {journal} {Prog. Part. Nucl. Phys.}\ }\textbf {\bibinfo
  {volume} {70}},\ \bibinfo {pages} {55} (\bibinfo {year} {2013})},\ \Eprint
  {http://arxiv.org/abs/1207.5999} {arXiv:1207.5999 [hep-lat]} \BibitemShut
  {NoStop}%
%%CITATION = ARXIV:1207.5999;%%
\bibitem [{\citenamefont {Borsanyi}(2017)}]{Borsanyi:2016bzg}%
  \BibitemOpen
  \bibfield  {author} {\bibinfo {author} {\bibfnamefont {S.}~\bibnamefont
  {Borsanyi}},\ }\bibfield  {booktitle} {\emph {\bibinfo {booktitle}
  {{Proceedings, 12th Conference on Quark Confinement and the Hadron Spectrum
  (Confinement XII): Thessaloniki, Greece}}},\ }\href {\doibase
  10.1051/epjconf/201713701006} {\bibfield  {journal} {\bibinfo  {journal} {EPJ
  Web Conf.}\ }\textbf {\bibinfo {volume} {137}},\ \bibinfo {pages} {01006}
  (\bibinfo {year} {2017})},\ \Eprint {http://arxiv.org/abs/1612.06755}
  {arXiv:1612.06755 [hep-lat]} \BibitemShut {NoStop}%
%%CITATION = ARXIV:1612.06755;%%
\bibitem [{\citenamefont {Barbour}\ \emph {et~al.}(1998)\citenamefont
  {Barbour}, \citenamefont {Morrison}, \citenamefont {Klepfish}, \citenamefont
  {Kogut},\ and\ \citenamefont {Lombardo}}]{Barbour:1997ej}%
  \BibitemOpen
  \bibfield  {author} {\bibinfo {author} {\bibfnamefont {I.~M.}\ \bibnamefont
  {Barbour}}, \bibinfo {author} {\bibfnamefont {S.~E.}\ \bibnamefont
  {Morrison}}, \bibinfo {author} {\bibfnamefont {E.~G.}\ \bibnamefont
  {Klepfish}}, \bibinfo {author} {\bibfnamefont {J.~B.}\ \bibnamefont {Kogut}},
  \ and\ \bibinfo {author} {\bibfnamefont {M.-P.}\ \bibnamefont {Lombardo}},\
  }\href {\doibase 10.1016/S0920-5632(97)00484-2} {\bibfield  {journal}
  {\bibinfo  {journal} {Nucl.\ Phys.\ B Proc.\ Suppl.}\ }\textbf {\bibinfo
  {volume} {60A}},\ \bibinfo {pages} {220} (\bibinfo {year} {1998})},\ \Eprint
  {http://arxiv.org/abs/hep-lat/9705042} {arXiv:hep-lat/9705042} \BibitemShut
  {NoStop}%
\bibitem [{\citenamefont {Fodor}\ and\ \citenamefont
  {Katz}(2002)}]{Fodor:2001pe}%
  \BibitemOpen
  \bibfield  {author} {\bibinfo {author} {\bibfnamefont {Z.}~\bibnamefont
  {Fodor}}\ and\ \bibinfo {author} {\bibfnamefont {S.~D.}\ \bibnamefont
  {Katz}},\ }\href {\doibase 10.1088/1126-6708/2002/03/014} {\bibfield
  {journal} {\bibinfo  {journal} {JHEP}\ }\textbf {\bibinfo {volume} {03}},\
  \bibinfo {pages} {014} (\bibinfo {year} {2002})},\ \Eprint
  {http://arxiv.org/abs/hep-lat/0106002} {arXiv:hep-lat/0106002 [hep-lat]}
  \BibitemShut {NoStop}%
%%CITATION = HEP-LAT/0106002;%%
\bibitem [{\citenamefont {De~Pietri}\ \emph {et~al.}(2007)\citenamefont
  {De~Pietri}, \citenamefont {Feo}, \citenamefont {Seiler},\ and\ \citenamefont
  {Stamatescu}}]{DePietri:2007ak}%
  \BibitemOpen
  \bibfield  {author} {\bibinfo {author} {\bibfnamefont {R.}~\bibnamefont
  {De~Pietri}}, \bibinfo {author} {\bibfnamefont {A.}~\bibnamefont {Feo}},
  \bibinfo {author} {\bibfnamefont {E.}~\bibnamefont {Seiler}}, \ and\ \bibinfo
  {author} {\bibfnamefont {I.-O.}\ \bibnamefont {Stamatescu}},\ }\href
  {\doibase 10.1103/PhysRevD.76.114501} {\bibfield  {journal} {\bibinfo
  {journal} {Phys. Rev.}\ }\textbf {\bibinfo {volume} {D76}},\ \bibinfo {pages}
  {114501} (\bibinfo {year} {2007})},\ \Eprint {http://arxiv.org/abs/0705.3420}
  {arXiv:0705.3420 [hep-lat]} \BibitemShut {NoStop}%
%%CITATION = ARXIV:0705.3420;%%
\bibitem [{\citenamefont {de~Forcrand}\ \emph {et~al.}(2000)\citenamefont
  {de~Forcrand} \emph {et~al.}}]{deForcrand:1999ih}%
  \BibitemOpen
  \bibfield  {author} {\bibinfo {author} {\bibfnamefont {P.}~\bibnamefont
  {de~Forcrand}} \emph {et~al.} (\bibinfo {collaboration} {QCD-TARO}),\ }\href
  {\doibase 10.1016/S0920-5632(00)00340-6} {\bibfield  {journal} {\bibinfo
  {journal} {Nucl.\ Phys.\ B Proc.\ Suppl.}\ }\textbf {\bibinfo {volume}
  {83}},\ \bibinfo {pages} {408} (\bibinfo {year} {2000})},\ \Eprint
  {http://arxiv.org/abs/hep-lat/9911034} {arXiv:hep-lat/9911034} \BibitemShut
  {NoStop}%
\bibitem [{\citenamefont {Miyamura}(2002)}]{Miyamura:2002mpl}%
  \BibitemOpen
  \bibfield  {author} {\bibinfo {author} {\bibfnamefont {O.}~\bibnamefont
  {Miyamura}} (\bibinfo {collaboration} {QCD-TARO}),\ }\href {\doibase
  10.1016/S0375-9474(01)01388-4} {\bibfield  {journal} {\bibinfo  {journal}
  {Nucl.\ Phys.\ A}\ }\textbf {\bibinfo {volume} {698}},\ \bibinfo {pages}
  {395} (\bibinfo {year} {2002})}\BibitemShut {NoStop}%
\bibitem [{\citenamefont {Kaczmarek}\ \emph {et~al.}(2011)\citenamefont
  {Kaczmarek}, \citenamefont {Karsch}, \citenamefont {Laermann}, \citenamefont
  {Miao}, \citenamefont {Mukherjee}, \citenamefont {Petreczky}, \citenamefont
  {Schmidt}, \citenamefont {Soeldner},\ and\ \citenamefont
  {Unger}}]{Kaczmarek:2011zz}%
  \BibitemOpen
  \bibfield  {author} {\bibinfo {author} {\bibfnamefont {O.}~\bibnamefont
  {Kaczmarek}}, \bibinfo {author} {\bibfnamefont {F.}~\bibnamefont {Karsch}},
  \bibinfo {author} {\bibfnamefont {E.}~\bibnamefont {Laermann}}, \bibinfo
  {author} {\bibfnamefont {C.}~\bibnamefont {Miao}}, \bibinfo {author}
  {\bibfnamefont {S.}~\bibnamefont {Mukherjee}}, \bibinfo {author}
  {\bibfnamefont {P.}~\bibnamefont {Petreczky}}, \bibinfo {author}
  {\bibfnamefont {C.}~\bibnamefont {Schmidt}}, \bibinfo {author} {\bibfnamefont
  {W.}~\bibnamefont {Soeldner}}, \ and\ \bibinfo {author} {\bibfnamefont
  {W.}~\bibnamefont {Unger}},\ }\href {\doibase 10.1103/PhysRevD.83.014504}
  {\bibfield  {journal} {\bibinfo  {journal} {Phys.\ Rev.\ D}\ }\textbf
  {\bibinfo {volume} {83}},\ \bibinfo {pages} {014504} (\bibinfo {year}
  {2011})},\ \Eprint {http://arxiv.org/abs/1011.3130} {arXiv:1011.3130
  [hep-lat]} \BibitemShut {NoStop}%
\bibitem [{\citenamefont {Endrodi}\ \emph {et~al.}(2011)\citenamefont
  {Endrodi}, \citenamefont {Fodor}, \citenamefont {Katz},\ and\ \citenamefont
  {Szabo}}]{Endrodi:2011gv}%
  \BibitemOpen
  \bibfield  {author} {\bibinfo {author} {\bibfnamefont {G.}~\bibnamefont
  {Endrodi}}, \bibinfo {author} {\bibfnamefont {Z.}~\bibnamefont {Fodor}},
  \bibinfo {author} {\bibfnamefont {S.}~\bibnamefont {Katz}}, \ and\ \bibinfo
  {author} {\bibfnamefont {K.}~\bibnamefont {Szabo}},\ }\href {\doibase
  10.1007/JHEP04(2011)001} {\bibfield  {journal} {\bibinfo  {journal} {JHEP}\
  }\textbf {\bibinfo {volume} {04}},\ \bibinfo {pages} {001} (\bibinfo {year}
  {2011})},\ \Eprint {http://arxiv.org/abs/1102.1356} {arXiv:1102.1356
  [hep-lat]} \BibitemShut {NoStop}%
\bibitem [{\citenamefont {Bonati}\ \emph {et~al.}(2018)\citenamefont {Bonati},
  \citenamefont {D'Elia}, \citenamefont {Negro}, \citenamefont {Sanfilippo},\
  and\ \citenamefont {Zambello}}]{Bonati:2018nut}%
  \BibitemOpen
  \bibfield  {author} {\bibinfo {author} {\bibfnamefont {C.}~\bibnamefont
  {Bonati}}, \bibinfo {author} {\bibfnamefont {M.}~\bibnamefont {D'Elia}},
  \bibinfo {author} {\bibfnamefont {F.}~\bibnamefont {Negro}}, \bibinfo
  {author} {\bibfnamefont {F.}~\bibnamefont {Sanfilippo}}, \ and\ \bibinfo
  {author} {\bibfnamefont {K.}~\bibnamefont {Zambello}},\ }\href {\doibase
  10.1103/PhysRevD.98.054510} {\bibfield  {journal} {\bibinfo  {journal}
  {Phys.\ Rev.\ D}\ }\textbf {\bibinfo {volume} {98}},\ \bibinfo {pages}
  {054510} (\bibinfo {year} {2018})},\ \Eprint
  {http://arxiv.org/abs/1805.02960} {arXiv:1805.02960 [hep-lat]} \BibitemShut
  {NoStop}%
\bibitem [{\citenamefont {Bazavov}\ \emph {et~al.}(2019)\citenamefont {Bazavov}
  \emph {et~al.}}]{Bazavov:2018mes}%
  \BibitemOpen
  \bibfield  {author} {\bibinfo {author} {\bibfnamefont {A.}~\bibnamefont
  {Bazavov}} \emph {et~al.} (\bibinfo {collaboration} {HotQCD}),\ }\href
  {\doibase 10.1016/j.physletb.2019.05.013} {\bibfield  {journal} {\bibinfo
  {journal} {Phys.\ Lett.\ B}\ }\textbf {\bibinfo {volume} {795}},\ \bibinfo
  {pages} {15} (\bibinfo {year} {2019})},\ \Eprint
  {http://arxiv.org/abs/1812.08235} {arXiv:1812.08235 [hep-lat]} \BibitemShut
  {NoStop}%
\bibitem [{\citenamefont {Cea}\ \emph {et~al.}(2014)\citenamefont {Cea},
  \citenamefont {Cosmai},\ and\ \citenamefont {Papa}}]{Cea:2014xva}%
  \BibitemOpen
  \bibfield  {author} {\bibinfo {author} {\bibfnamefont {P.}~\bibnamefont
  {Cea}}, \bibinfo {author} {\bibfnamefont {L.}~\bibnamefont {Cosmai}}, \ and\
  \bibinfo {author} {\bibfnamefont {A.}~\bibnamefont {Papa}},\ }\href {\doibase
  10.1103/PhysRevD.89.074512} {\bibfield  {journal} {\bibinfo  {journal}
  {Phys.\ Rev.\ D}\ }\textbf {\bibinfo {volume} {89}},\ \bibinfo {pages}
  {074512} (\bibinfo {year} {2014})},\ \Eprint {http://arxiv.org/abs/1403.0821}
  {arXiv:1403.0821 [hep-lat]} \BibitemShut {NoStop}%
\bibitem [{\citenamefont {Bonati}\ \emph {et~al.}(2015)\citenamefont {Bonati},
  \citenamefont {D'Elia}, \citenamefont {Mariti}, \citenamefont {Mesiti},
  \citenamefont {Negro},\ and\ \citenamefont {Sanfilippo}}]{Bonati:2015bha}%
  \BibitemOpen
  \bibfield  {author} {\bibinfo {author} {\bibfnamefont {C.}~\bibnamefont
  {Bonati}}, \bibinfo {author} {\bibfnamefont {M.}~\bibnamefont {D'Elia}},
  \bibinfo {author} {\bibfnamefont {M.}~\bibnamefont {Mariti}}, \bibinfo
  {author} {\bibfnamefont {M.}~\bibnamefont {Mesiti}}, \bibinfo {author}
  {\bibfnamefont {F.}~\bibnamefont {Negro}}, \ and\ \bibinfo {author}
  {\bibfnamefont {F.}~\bibnamefont {Sanfilippo}},\ }\href {\doibase
  10.1103/PhysRevD.92.054503} {\bibfield  {journal} {\bibinfo  {journal}
  {Phys.\ Rev.\ D}\ }\textbf {\bibinfo {volume} {92}},\ \bibinfo {pages}
  {054503} (\bibinfo {year} {2015})},\ \Eprint
  {http://arxiv.org/abs/1507.03571} {arXiv:1507.03571 [hep-lat]} \BibitemShut
  {NoStop}%
\bibitem [{\citenamefont {Bellwied}\ \emph {et~al.}(2015)\citenamefont
  {Bellwied}, \citenamefont {Borsanyi}, \citenamefont {Fodor}, \citenamefont
  {Günther}, \citenamefont {Katz}, \citenamefont {Ratti},\ and\ \citenamefont
  {Szabo}}]{Bellwied:2015rza}%
  \BibitemOpen
  \bibfield  {author} {\bibinfo {author} {\bibfnamefont {R.}~\bibnamefont
  {Bellwied}}, \bibinfo {author} {\bibfnamefont {S.}~\bibnamefont {Borsanyi}},
  \bibinfo {author} {\bibfnamefont {Z.}~\bibnamefont {Fodor}}, \bibinfo
  {author} {\bibfnamefont {J.}~\bibnamefont {Günther}}, \bibinfo {author}
  {\bibfnamefont {S.}~\bibnamefont {Katz}}, \bibinfo {author} {\bibfnamefont
  {C.}~\bibnamefont {Ratti}}, \ and\ \bibinfo {author} {\bibfnamefont
  {K.}~\bibnamefont {Szabo}},\ }\href {\doibase 10.1016/j.physletb.2015.11.011}
  {\bibfield  {journal} {\bibinfo  {journal} {Phys.\ Lett.\ B}\ }\textbf
  {\bibinfo {volume} {751}},\ \bibinfo {pages} {559} (\bibinfo {year}
  {2015})},\ \Eprint {http://arxiv.org/abs/1507.07510} {arXiv:1507.07510
  [hep-lat]} \BibitemShut {NoStop}%
\bibitem [{\citenamefont {Borsanyi}\ \emph {et~al.}(2020)\citenamefont
  {Borsanyi}, \citenamefont {Fodor}, \citenamefont {Guenther}, \citenamefont
  {Kara}, \citenamefont {Katz}, \citenamefont {Parotto}, \citenamefont
  {Pasztor}, \citenamefont {Ratti},\ and\ \citenamefont
  {Szabo}}]{Borsanyi:2020fev}%
  \BibitemOpen
  \bibfield  {author} {\bibinfo {author} {\bibfnamefont {S.}~\bibnamefont
  {Borsanyi}}, \bibinfo {author} {\bibfnamefont {Z.}~\bibnamefont {Fodor}},
  \bibinfo {author} {\bibfnamefont {J.~N.}\ \bibnamefont {Guenther}}, \bibinfo
  {author} {\bibfnamefont {R.}~\bibnamefont {Kara}}, \bibinfo {author}
  {\bibfnamefont {S.~D.}\ \bibnamefont {Katz}}, \bibinfo {author}
  {\bibfnamefont {P.}~\bibnamefont {Parotto}}, \bibinfo {author} {\bibfnamefont
  {A.}~\bibnamefont {Pasztor}}, \bibinfo {author} {\bibfnamefont
  {C.}~\bibnamefont {Ratti}}, \ and\ \bibinfo {author} {\bibfnamefont {K.~K.}\
  \bibnamefont {Szabo}},\ }\href@noop {} {\  (\bibinfo {year} {2020})},\
  \Eprint {http://arxiv.org/abs/2002.02821} {arXiv:2002.02821 [hep-lat]}
  \BibitemShut {NoStop}%
\bibitem [{\citenamefont {Parisi}(1983)}]{Parisi:1984cs}%
  \BibitemOpen
  \bibfield  {author} {\bibinfo {author} {\bibfnamefont {G.}~\bibnamefont
  {Parisi}},\ }\href {\doibase 10.1016/0370-2693(83)90525-7} {\bibfield
  {journal} {\bibinfo  {journal} {Phys.Lett.}\ }\textbf {\bibinfo {volume}
  {B131}},\ \bibinfo {pages} {393} (\bibinfo {year} {1983})}\BibitemShut
  {NoStop}%
%%CITATION = PHLTA,B131,393;%%
\bibitem [{\citenamefont {Klauder}(1983)}]{Klauder:1983nn}%
  \BibitemOpen
  \bibfield  {author} {\bibinfo {author} {\bibfnamefont {J.~R.}\ \bibnamefont
  {Klauder}},\ }\href {\doibase 10.1007/978-3-7091-7651-1_8} {\bibfield
  {journal} {\bibinfo  {journal} {Acta Phys.Austriaca Suppl.}\ }\textbf
  {\bibinfo {volume} {25}},\ \bibinfo {pages} {251} (\bibinfo {year}
  {1983})}\BibitemShut {NoStop}%
%%CITATION = APAUA,25,251;%%
\bibitem [{\citenamefont {Berges}\ and\ \citenamefont
  {Stamatescu}(2005)}]{Berges2005b}%
  \BibitemOpen
  \bibfield  {author} {\bibinfo {author} {\bibfnamefont {J.}~\bibnamefont
  {Berges}}\ and\ \bibinfo {author} {\bibfnamefont {I.-O.}\ \bibnamefont
  {Stamatescu}},\ }\href@noop {} {\bibfield  {journal} {\bibinfo  {journal}
  {Phys. Rev. Lett.}\ }\textbf {\bibinfo {volume} {95}},\ \bibinfo {pages}
  {202003} (\bibinfo {year} {2005})}\BibitemShut {NoStop}%
\bibitem [{\citenamefont {Aarts}\ and\ \citenamefont
  {Stamatescu}(2008)}]{Aarts:2008rr}%
  \BibitemOpen
  \bibfield  {author} {\bibinfo {author} {\bibfnamefont {G.}~\bibnamefont
  {Aarts}}\ and\ \bibinfo {author} {\bibfnamefont {I.-O.}\ \bibnamefont
  {Stamatescu}},\ }\href {\doibase 10.1088/1126-6708/2008/09/018} {\bibfield
  {journal} {\bibinfo  {journal} {JHEP}\ }\textbf {\bibinfo {volume} {0809}},\
  \bibinfo {pages} {018} (\bibinfo {year} {2008})},\ \Eprint
  {http://arxiv.org/abs/0807.1597} {arXiv:0807.1597 [hep-lat]} \BibitemShut
  {NoStop}%
%%CITATION = ARXIV:0807.1597;%%
\bibitem [{\citenamefont {Berger}\ \emph {et~al.}(2019)\citenamefont {Berger},
  \citenamefont {Rammelmüller}, \citenamefont {Loheac}, \citenamefont
  {Ehmann}, \citenamefont {Braun},\ and\ \citenamefont
  {Drut}}]{Berger:2019odf}%
  \BibitemOpen
  \bibfield  {author} {\bibinfo {author} {\bibfnamefont {C.~E.}\ \bibnamefont
  {Berger}}, \bibinfo {author} {\bibfnamefont {L.}~\bibnamefont
  {Rammelmüller}}, \bibinfo {author} {\bibfnamefont {A.~C.}\ \bibnamefont
  {Loheac}}, \bibinfo {author} {\bibfnamefont {F.}~\bibnamefont {Ehmann}},
  \bibinfo {author} {\bibfnamefont {J.}~\bibnamefont {Braun}}, \ and\ \bibinfo
  {author} {\bibfnamefont {J.~E.}\ \bibnamefont {Drut}},\ }\href@noop {} {\
  (\bibinfo {year} {2019})},\ \Eprint {http://arxiv.org/abs/1907.10183}
  {arXiv:1907.10183 [cond-mat.quant-gas]} \BibitemShut {NoStop}%
\bibitem [{\citenamefont {Attanasio}\ \emph {et~al.}(2020)\citenamefont
  {Attanasio}, \citenamefont {Jäger},\ and\ \citenamefont
  {Ziegler}}]{Attanasio:2020spv}%
  \BibitemOpen
  \bibfield  {author} {\bibinfo {author} {\bibfnamefont {F.}~\bibnamefont
  {Attanasio}}, \bibinfo {author} {\bibfnamefont {B.}~\bibnamefont {Jäger}}, \
  and\ \bibinfo {author} {\bibfnamefont {F.~P.}\ \bibnamefont {Ziegler}},\
  }\href@noop {} {\  (\bibinfo {year} {2020})},\ \Eprint
  {http://arxiv.org/abs/2006.00476} {arXiv:2006.00476 [hep-lat]} \BibitemShut
  {NoStop}%
\bibitem [{\citenamefont {Batrouni}\ \emph {et~al.}(1985)\citenamefont
  {Batrouni}, \citenamefont {Katz}, \citenamefont {Kronfeld}, \citenamefont
  {Lepage}, \citenamefont {Svetitsky},\ and\ \citenamefont
  {Wilson}}]{PhysRevD.32.2736}%
  \BibitemOpen
  \bibfield  {author} {\bibinfo {author} {\bibfnamefont {G.~G.}\ \bibnamefont
  {Batrouni}}, \bibinfo {author} {\bibfnamefont {G.~R.}\ \bibnamefont {Katz}},
  \bibinfo {author} {\bibfnamefont {A.~S.}\ \bibnamefont {Kronfeld}}, \bibinfo
  {author} {\bibfnamefont {G.~P.}\ \bibnamefont {Lepage}}, \bibinfo {author}
  {\bibfnamefont {B.}~\bibnamefont {Svetitsky}}, \ and\ \bibinfo {author}
  {\bibfnamefont {K.~G.}\ \bibnamefont {Wilson}},\ }\href {\doibase
  10.1103/PhysRevD.32.2736} {\bibfield  {journal} {\bibinfo  {journal} {Phys.
  Rev. D}\ }\textbf {\bibinfo {volume} {32}},\ \bibinfo {pages} {2736}
  (\bibinfo {year} {1985})}\BibitemShut {NoStop}%
\bibitem [{\citenamefont {Mollgaard}\ and\ \citenamefont
  {Splittorff}(2013)}]{Mollgaard:2013qra}%
  \BibitemOpen
  \bibfield  {author} {\bibinfo {author} {\bibfnamefont {A.}~\bibnamefont
  {Mollgaard}}\ and\ \bibinfo {author} {\bibfnamefont {K.}~\bibnamefont
  {Splittorff}},\ }\href {\doibase 10.1103/PhysRevD.88.116007} {\bibfield
  {journal} {\bibinfo  {journal} {Phys.Rev.}\ }\textbf {\bibinfo {volume}
  {D88}},\ \bibinfo {pages} {116007} (\bibinfo {year} {2013})},\ \Eprint
  {http://arxiv.org/abs/1309.4335} {arXiv:1309.4335 [hep-lat]} \BibitemShut
  {NoStop}%
%%CITATION = ARXIV:1309.4335;%%
\bibitem [{\citenamefont {Greensite}(2014)}]{Greensite:2014cxa}%
  \BibitemOpen
  \bibfield  {author} {\bibinfo {author} {\bibfnamefont {J.}~\bibnamefont
  {Greensite}},\ }\href {\doibase 10.1103/PhysRevD.90.114507} {\bibfield
  {journal} {\bibinfo  {journal} {Phys.Rev.}\ }\textbf {\bibinfo {volume}
  {D90}},\ \bibinfo {pages} {114507} (\bibinfo {year} {2014})},\ \Eprint
  {http://arxiv.org/abs/1406.4558} {arXiv:1406.4558 [hep-lat]} \BibitemShut
  {NoStop}%
%%CITATION = ARXIV:1406.4558;%%
\bibitem [{\citenamefont {Nishimura}\ and\ \citenamefont
  {Shimasaki}(2015)}]{Nishimura:2015pba}%
  \BibitemOpen
  \bibfield  {author} {\bibinfo {author} {\bibfnamefont {J.}~\bibnamefont
  {Nishimura}}\ and\ \bibinfo {author} {\bibfnamefont {S.}~\bibnamefont
  {Shimasaki}},\ }\href {\doibase 10.1103/PhysRevD.92.011501} {\bibfield
  {journal} {\bibinfo  {journal} {Phys. Rev.}\ }\textbf {\bibinfo {volume}
  {D92}},\ \bibinfo {pages} {011501} (\bibinfo {year} {2015})},\ \Eprint
  {http://arxiv.org/abs/1504.08359} {arXiv:1504.08359 [hep-lat]} \BibitemShut
  {NoStop}%
%%CITATION = ARXIV:1504.08359;%%
\bibitem [{\citenamefont {Aarts}\ \emph {et~al.}(2017)\citenamefont {Aarts},
  \citenamefont {Seiler}, \citenamefont {Sexty},\ and\ \citenamefont
  {Stamatescu}}]{Aarts:2017vrv}%
  \BibitemOpen
  \bibfield  {author} {\bibinfo {author} {\bibfnamefont {G.}~\bibnamefont
  {Aarts}}, \bibinfo {author} {\bibfnamefont {E.}~\bibnamefont {Seiler}},
  \bibinfo {author} {\bibfnamefont {D.}~\bibnamefont {Sexty}}, \ and\ \bibinfo
  {author} {\bibfnamefont {I.-O.}\ \bibnamefont {Stamatescu}},\ }\href
  {\doibase 10.1007/JHEP05(2017)044, 10.1007/JHEP01(2018)128} {\bibfield
  {journal} {\bibinfo  {journal} {JHEP}\ }\textbf {\bibinfo {volume} {05}},\
  \bibinfo {pages} {044} (\bibinfo {year} {2017})},\ \bibinfo {note} {[Erratum:
  JHEP01,128(2018)]},\ \Eprint {http://arxiv.org/abs/1701.02322}
  {arXiv:1701.02322 [hep-lat]} \BibitemShut {NoStop}%
%%CITATION = ARXIV:1701.02322;%%
\bibitem [{\citenamefont {Seiler}\ \emph {et~al.}(2013)\citenamefont {Seiler},
  \citenamefont {Sexty},\ and\ \citenamefont {Stamatescu}}]{gaugecooling}%
  \BibitemOpen
  \bibfield  {author} {\bibinfo {author} {\bibfnamefont {E.}~\bibnamefont
  {Seiler}}, \bibinfo {author} {\bibfnamefont {D.}~\bibnamefont {Sexty}}, \
  and\ \bibinfo {author} {\bibfnamefont {I.-O.}\ \bibnamefont {Stamatescu}},\
  }\href {\doibase 10.1016/j.physletb.2013.04.062} {\bibfield  {journal}
  {\bibinfo  {journal} {Phys.Lett.}\ }\textbf {\bibinfo {volume} {B723}},\
  \bibinfo {pages} {213} (\bibinfo {year} {2013})},\ \Eprint
  {http://arxiv.org/abs/1211.3709} {arXiv:1211.3709 [hep-lat]} \BibitemShut
  {NoStop}%
%%CITATION = ARXIV:1211.3709;%%
\bibitem [{\citenamefont {Aarts}\ \emph {et~al.}(2013)\citenamefont {Aarts},
  \citenamefont {Bongiovanni}, \citenamefont {Seiler}, \citenamefont {Sexty},\
  and\ \citenamefont {Stamatescu}}]{Aarts:2013uxa}%
  \BibitemOpen
  \bibfield  {author} {\bibinfo {author} {\bibfnamefont {G.}~\bibnamefont
  {Aarts}}, \bibinfo {author} {\bibfnamefont {L.}~\bibnamefont {Bongiovanni}},
  \bibinfo {author} {\bibfnamefont {E.}~\bibnamefont {Seiler}}, \bibinfo
  {author} {\bibfnamefont {D.}~\bibnamefont {Sexty}}, \ and\ \bibinfo {author}
  {\bibfnamefont {I.-O.}\ \bibnamefont {Stamatescu}},\ }\href {\doibase
  10.1140/epja/i2013-13089-4} {\bibfield  {journal} {\bibinfo  {journal}
  {Eur.Phys.J.}\ }\textbf {\bibinfo {volume} {A49}},\ \bibinfo {pages} {89}
  (\bibinfo {year} {2013})},\ \Eprint {http://arxiv.org/abs/1303.6425}
  {arXiv:1303.6425 [hep-lat]} \BibitemShut {NoStop}%
%%CITATION = ARXIV:1303.6425;%%
\bibitem [{\citenamefont {Aarts}\ \emph {et~al.}(2016)\citenamefont {Aarts},
  \citenamefont {Attanasio}, \citenamefont {Jäger},\ and\ \citenamefont
  {Sexty}}]{hdqcdpd}%
  \BibitemOpen
  \bibfield  {author} {\bibinfo {author} {\bibfnamefont {G.}~\bibnamefont
  {Aarts}}, \bibinfo {author} {\bibfnamefont {F.}~\bibnamefont {Attanasio}},
  \bibinfo {author} {\bibfnamefont {B.}~\bibnamefont {Jäger}}, \ and\ \bibinfo
  {author} {\bibfnamefont {D.}~\bibnamefont {Sexty}},\ }\href {\doibase
  10.1007/JHEP09(2016)087} {\bibfield  {journal} {\bibinfo  {journal} {JHEP}\
  }\textbf {\bibinfo {volume} {09}},\ \bibinfo {pages} {087} (\bibinfo {year}
  {2016})},\ \Eprint {http://arxiv.org/abs/1606.05561} {arXiv:1606.05561
  [hep-lat]} \BibitemShut {NoStop}%
%%CITATION = ARXIV:1606.05561;%%
\bibitem [{\citenamefont {Scherzer}\ \emph
  {et~al.}(2019{\natexlab{a}})\citenamefont {Scherzer}, \citenamefont {Seiler},
  \citenamefont {Sexty},\ and\ \citenamefont {Stamatescu}}]{boundaryterms1}%
  \BibitemOpen
  \bibfield  {author} {\bibinfo {author} {\bibfnamefont {M.}~\bibnamefont
  {Scherzer}}, \bibinfo {author} {\bibfnamefont {E.}~\bibnamefont {Seiler}},
  \bibinfo {author} {\bibfnamefont {D.}~\bibnamefont {Sexty}}, \ and\ \bibinfo
  {author} {\bibfnamefont {I.-O.}\ \bibnamefont {Stamatescu}},\ }\href
  {\doibase 10.1103/PhysRevD.99.014512} {\bibfield  {journal} {\bibinfo
  {journal} {Phys. Rev.}\ }\textbf {\bibinfo {volume} {D99}},\ \bibinfo {pages}
  {014512} (\bibinfo {year} {2019}{\natexlab{a}})},\ \Eprint
  {http://arxiv.org/abs/1808.05187} {arXiv:1808.05187 [hep-lat]} \BibitemShut
  {NoStop}%
%%CITATION = ARXIV:1808.05187;%%
\bibitem [{\citenamefont {Berges}\ \emph {et~al.}(2007)\citenamefont {Berges},
  \citenamefont {Borsanyi}, \citenamefont {Sexty},\ and\ \citenamefont
  {Stamatescu}}]{Berges:2006xc}%
  \BibitemOpen
  \bibfield  {author} {\bibinfo {author} {\bibfnamefont {J.}~\bibnamefont
  {Berges}}, \bibinfo {author} {\bibfnamefont {S.}~\bibnamefont {Borsanyi}},
  \bibinfo {author} {\bibfnamefont {D.}~\bibnamefont {Sexty}}, \ and\ \bibinfo
  {author} {\bibfnamefont {I.~O.}\ \bibnamefont {Stamatescu}},\ }\href
  {\doibase 10.1103/PhysRevD.75.045007} {\bibfield  {journal} {\bibinfo
  {journal} {Phys. Rev. D}\ }\textbf {\bibinfo {volume} {75}},\ \bibinfo
  {pages} {045007} (\bibinfo {year} {2007})},\ \Eprint
  {http://arxiv.org/abs/hep-lat/0609058} {arXiv:hep-lat/0609058} \BibitemShut
  {NoStop}%
%%CITATION = HEP-LAT/0609058;%%
\bibitem [{\citenamefont {Berges}\ and\ \citenamefont
  {Sexty}(2008)}]{Berges:2007nr}%
  \BibitemOpen
  \bibfield  {author} {\bibinfo {author} {\bibfnamefont {J.}~\bibnamefont
  {Berges}}\ and\ \bibinfo {author} {\bibfnamefont {D.}~\bibnamefont {Sexty}},\
  }\href {\doibase 10.1016/j.nuclphysb.2008.01.018} {\bibfield  {journal}
  {\bibinfo  {journal} {Nucl. Phys.}\ }\textbf {\bibinfo {volume} {B799}},\
  \bibinfo {pages} {306} (\bibinfo {year} {2008})},\ \Eprint
  {http://arxiv.org/abs/0708.0779} {arXiv:0708.0779 [hep-lat]} \BibitemShut
  {NoStop}%
%%CITATION = 0708.0779;%%
\bibitem [{\citenamefont {Bender}\ \emph {et~al.}(1992)\citenamefont {Bender},
  \citenamefont {Hashimoto}, \citenamefont {Karsch}, \citenamefont {Linke},
  \citenamefont {Nakamura}, \citenamefont {Plewnia}, \citenamefont
  {Stamatescu},\ and\ \citenamefont {Wetzel}}]{Bender:1992gn}%
  \BibitemOpen
  \bibfield  {author} {\bibinfo {author} {\bibfnamefont {I.}~\bibnamefont
  {Bender}}, \bibinfo {author} {\bibfnamefont {T.}~\bibnamefont {Hashimoto}},
  \bibinfo {author} {\bibfnamefont {F.}~\bibnamefont {Karsch}}, \bibinfo
  {author} {\bibfnamefont {V.}~\bibnamefont {Linke}}, \bibinfo {author}
  {\bibfnamefont {A.}~\bibnamefont {Nakamura}}, \bibinfo {author}
  {\bibfnamefont {M.}~\bibnamefont {Plewnia}}, \bibinfo {author} {\bibfnamefont
  {I.~O.}\ \bibnamefont {Stamatescu}}, \ and\ \bibinfo {author} {\bibfnamefont
  {W.}~\bibnamefont {Wetzel}},\ }\bibfield  {booktitle} {\emph {\bibinfo
  {booktitle} {{LATTICE 91: International Symposium on Lattice Field Theory
  Tsukuba, Japan, November 5-9, 1991}}},\ }\href {\doibase
  10.1016/0920-5632(92)90265-T} {\bibfield  {journal} {\bibinfo  {journal}
  {Nucl. Phys. Proc. Suppl.}\ }\textbf {\bibinfo {volume} {26}},\ \bibinfo
  {pages} {323} (\bibinfo {year} {1992})}\BibitemShut {NoStop}%
%%CITATION = NUPHZ,26,323;%%
\bibitem [{\citenamefont {Blum}\ \emph {et~al.}(1996)\citenamefont {Blum},
  \citenamefont {Hetrick},\ and\ \citenamefont {Toussaint}}]{Blum:1995cb}%
  \BibitemOpen
  \bibfield  {author} {\bibinfo {author} {\bibfnamefont {T.~C.}\ \bibnamefont
  {Blum}}, \bibinfo {author} {\bibfnamefont {J.~E.}\ \bibnamefont {Hetrick}}, \
  and\ \bibinfo {author} {\bibfnamefont {D.}~\bibnamefont {Toussaint}},\ }\href
  {\doibase 10.1103/PhysRevLett.76.1019} {\bibfield  {journal} {\bibinfo
  {journal} {Phys. Rev. Lett.}\ }\textbf {\bibinfo {volume} {76}},\ \bibinfo
  {pages} {1019} (\bibinfo {year} {1996})},\ \Eprint
  {http://arxiv.org/abs/hep-lat/9509002} {arXiv:hep-lat/9509002 [hep-lat]}
  \BibitemShut {NoStop}%
%%CITATION = HEP-LAT/9509002;%%
\bibitem [{\citenamefont {Scherzer}\ \emph
  {et~al.}(2019{\natexlab{b}})\citenamefont {Scherzer}, \citenamefont {Seiler},
  \citenamefont {Sexty},\ and\ \citenamefont {Stamatescu}}]{boundaryterms2}%
  \BibitemOpen
  \bibfield  {author} {\bibinfo {author} {\bibfnamefont {M.}~\bibnamefont
  {Scherzer}}, \bibinfo {author} {\bibfnamefont {E.}~\bibnamefont {Seiler}},
  \bibinfo {author} {\bibfnamefont {D.}~\bibnamefont {Sexty}}, \ and\ \bibinfo
  {author} {\bibfnamefont {I.~O.}\ \bibnamefont {Stamatescu}},\ }\href@noop {}
  {\  (\bibinfo {year} {2019}{\natexlab{b}})},\ \Eprint
  {http://arxiv.org/abs/1910.09427} {arXiv:1910.09427 [hep-lat]} \BibitemShut
  {NoStop}%
%%CITATION = ARXIV:1910.09427;%%
\bibitem [{\citenamefont {Nagata}\ \emph {et~al.}(2016)\citenamefont {Nagata},
  \citenamefont {Nishimura},\ and\ \citenamefont {Shimasaki}}]{Nagata:2016vkn}%
  \BibitemOpen
  \bibfield  {author} {\bibinfo {author} {\bibfnamefont {K.}~\bibnamefont
  {Nagata}}, \bibinfo {author} {\bibfnamefont {J.}~\bibnamefont {Nishimura}}, \
  and\ \bibinfo {author} {\bibfnamefont {S.}~\bibnamefont {Shimasaki}},\ }\href
  {\doibase 10.1103/PhysRevD.94.114515} {\bibfield  {journal} {\bibinfo
  {journal} {Phys.\ Rev.\ D}\ }\textbf {\bibinfo {volume} {94}},\ \bibinfo
  {pages} {114515} (\bibinfo {year} {2016})},\ \Eprint
  {http://arxiv.org/abs/1606.07627} {arXiv:1606.07627 [hep-lat]} \BibitemShut
  {NoStop}%
\bibitem [{\citenamefont {Fodor}\ \emph {et~al.}(2015)\citenamefont {Fodor},
  \citenamefont {Katz}, \citenamefont {Sexty},\ and\ \citenamefont
  {T{\"o}r{\"o}k}}]{Fodor:2015doa}%
  \BibitemOpen
  \bibfield  {author} {\bibinfo {author} {\bibfnamefont {Z.}~\bibnamefont
  {Fodor}}, \bibinfo {author} {\bibfnamefont {S.~D.}\ \bibnamefont {Katz}},
  \bibinfo {author} {\bibfnamefont {D.}~\bibnamefont {Sexty}}, \ and\ \bibinfo
  {author} {\bibfnamefont {C.}~\bibnamefont {T{\"o}r{\"o}k}},\ }\href {\doibase
  10.1103/PhysRevD.92.094516} {\bibfield  {journal} {\bibinfo  {journal} {Phys.
  Rev.}\ }\textbf {\bibinfo {volume} {D92}},\ \bibinfo {pages} {094516}
  (\bibinfo {year} {2015})},\ \Eprint {http://arxiv.org/abs/1508.05260}
  {arXiv:1508.05260 [hep-lat]} \BibitemShut {NoStop}%
%%CITATION = ARXIV:1508.05260;%%
\bibitem [{\citenamefont {Sexty}(2019)}]{Sexty:2019vqx}%
  \BibitemOpen
  \bibfield  {author} {\bibinfo {author} {\bibfnamefont {D.}~\bibnamefont
  {Sexty}},\ }\href {\doibase 10.1103/PhysRevD.100.074503} {\bibfield
  {journal} {\bibinfo  {journal} {Phys. Rev. D}\ }\textbf {\bibinfo {volume}
  {100}},\ \bibinfo {pages} {074503} (\bibinfo {year} {2019})}\BibitemShut
  {NoStop}%
\bibitem [{\citenamefont {Sexty}(2014)}]{Sexty:2013ica}%
  \BibitemOpen
  \bibfield  {author} {\bibinfo {author} {\bibfnamefont {D.}~\bibnamefont
  {Sexty}},\ }\href {\doibase 10.1016/j.physletb.2014.01.019} {\bibfield
  {journal} {\bibinfo  {journal} {Phys.Lett.}\ }\textbf {\bibinfo {volume}
  {B729}},\ \bibinfo {pages} {108} (\bibinfo {year} {2014})},\ \Eprint
  {http://arxiv.org/abs/1307.7748} {arXiv:1307.7748 [hep-lat]} \BibitemShut
  {NoStop}%
%%CITATION = ARXIV:1307.7748;%%
\bibitem [{\citenamefont {Aarts}\ \emph {et~al.}(2010)\citenamefont {Aarts},
  \citenamefont {James}, \citenamefont {Seiler},\ and\ \citenamefont
  {Stamatescu}}]{Aarts:2009dg}%
  \BibitemOpen
  \bibfield  {author} {\bibinfo {author} {\bibfnamefont {G.}~\bibnamefont
  {Aarts}}, \bibinfo {author} {\bibfnamefont {F.~A.}\ \bibnamefont {James}},
  \bibinfo {author} {\bibfnamefont {E.}~\bibnamefont {Seiler}}, \ and\ \bibinfo
  {author} {\bibfnamefont {I.-O.}\ \bibnamefont {Stamatescu}},\ }\href
  {\doibase 10.1016/j.physletb.2010.03.012} {\bibfield  {journal} {\bibinfo
  {journal} {Phys.Lett.}\ }\textbf {\bibinfo {volume} {B687}},\ \bibinfo
  {pages} {154} (\bibinfo {year} {2010})},\ \Eprint
  {http://arxiv.org/abs/0912.0617} {arXiv:0912.0617 [hep-lat]} \BibitemShut
  {NoStop}%
%%CITATION = ARXIV:0912.0617;%%
\bibitem [{\citenamefont {Borsanyi}\ \emph {et~al.}(2012)\citenamefont
  {Borsanyi}, \citenamefont {Durr}, \citenamefont {Fodor}, \citenamefont
  {Hoelbling}, \citenamefont {Katz} \emph {et~al.}}]{Borsanyi:2012zs}%
  \BibitemOpen
  \bibfield  {author} {\bibinfo {author} {\bibfnamefont {S.}~\bibnamefont
  {Borsanyi}}, \bibinfo {author} {\bibfnamefont {S.}~\bibnamefont {Durr}},
  \bibinfo {author} {\bibfnamefont {Z.}~\bibnamefont {Fodor}}, \bibinfo
  {author} {\bibfnamefont {C.}~\bibnamefont {Hoelbling}}, \bibinfo {author}
  {\bibfnamefont {S.~D.}\ \bibnamefont {Katz}},  \emph {et~al.},\ }\href
  {\doibase 10.1007/JHEP09(2012)010} {\bibfield  {journal} {\bibinfo  {journal}
  {JHEP}\ }\textbf {\bibinfo {volume} {1209}},\ \bibinfo {pages} {010}
  (\bibinfo {year} {2012})},\ \Eprint {http://arxiv.org/abs/1203.4469}
  {arXiv:1203.4469 [hep-lat]} \BibitemShut {NoStop}%
%%CITATION = ARXIV:1203.4469;%%
\bibitem [{\citenamefont {Splittorff}(2015)}]{Splittorff:2014zca}%
  \BibitemOpen
  \bibfield  {author} {\bibinfo {author} {\bibfnamefont {K.}~\bibnamefont
  {Splittorff}},\ }\href {\doibase 10.1103/PhysRevD.91.034507} {\bibfield
  {journal} {\bibinfo  {journal} {Phys. Rev. D}\ }\textbf {\bibinfo {volume}
  {91}},\ \bibinfo {pages} {034507} (\bibinfo {year} {2015})},\ \Eprint
  {http://arxiv.org/abs/1412.0502} {arXiv:1412.0502 [hep-lat]} \BibitemShut
  {NoStop}%
\bibitem [{\citenamefont {Bloch}\ and\ \citenamefont
  {Schenk}(2018)}]{Bloch:2017jzi}%
  \BibitemOpen
  \bibfield  {author} {\bibinfo {author} {\bibfnamefont {J.}~\bibnamefont
  {Bloch}}\ and\ \bibinfo {author} {\bibfnamefont {O.}~\bibnamefont {Schenk}},\
  }\bibfield  {booktitle} {\emph {\bibinfo {booktitle} {{Proceedings, 35th
  International Symposium on Lattice Field Theory (Lattice 2017): Granada,
  Spain, June 18-24, 2017}}},\ }\href {\doibase 10.1051/epjconf/201817507003}
  {\bibfield  {journal} {\bibinfo  {journal} {EPJ Web Conf.}\ }\textbf
  {\bibinfo {volume} {175}},\ \bibinfo {pages} {07003} (\bibinfo {year}
  {2018})},\ \Eprint {http://arxiv.org/abs/1707.08874} {arXiv:1707.08874
  [hep-lat]} \BibitemShut {NoStop}%
%%CITATION = ARXIV:1707.08874;%%
\bibitem [{\citenamefont {Kogut}\ and\ \citenamefont
  {Sinclair}(2019)}]{Kogut:2019qmi}%
  \BibitemOpen
  \bibfield  {author} {\bibinfo {author} {\bibfnamefont {J.~B.}\ \bibnamefont
  {Kogut}}\ and\ \bibinfo {author} {\bibfnamefont {D.~K.}\ \bibnamefont
  {Sinclair}},\ }\href {\doibase 10.1103/PhysRevD.100.054512} {\bibfield
  {journal} {\bibinfo  {journal} {Phys. Rev.}\ }\textbf {\bibinfo {volume}
  {D100}},\ \bibinfo {pages} {054512} (\bibinfo {year} {2019})},\ \Eprint
  {http://arxiv.org/abs/1903.02622} {arXiv:1903.02622 [hep-lat]} \BibitemShut
  {NoStop}%
%%CITATION = ARXIV:1903.02622;%%
\bibitem [{\citenamefont {Binder}(1981)}]{binder1981finite}%
  \BibitemOpen
  \bibfield  {author} {\bibinfo {author} {\bibfnamefont {K.}~\bibnamefont
  {Binder}},\ }\href@noop {} {\bibfield  {journal} {\bibinfo  {journal}
  {Zeitschrift f{\"u}r Physik B Condensed Matter}\ }\textbf {\bibinfo {volume}
  {43}},\ \bibinfo {pages} {119} (\bibinfo {year} {1981})}\BibitemShut
  {NoStop}%
\bibitem [{\citenamefont {Chen}\ \emph {et~al.}(2010)\citenamefont {Chen},
  \citenamefont {Pan}, \citenamefont {Chen},\ and\ \citenamefont
  {Wu}}]{Chen:2010ej}%
  \BibitemOpen
  \bibfield  {author} {\bibinfo {author} {\bibfnamefont {L.}~\bibnamefont
  {Chen}}, \bibinfo {author} {\bibfnamefont {X.}~\bibnamefont {Pan}}, \bibinfo
  {author} {\bibfnamefont {X.}~\bibnamefont {Chen}}, \ and\ \bibinfo {author}
  {\bibfnamefont {Y.}~\bibnamefont {Wu}},\ }\href@noop {} {\  (\bibinfo {year}
  {2010})},\ \Eprint {http://arxiv.org/abs/1010.1166} {arXiv:1010.1166
  [nucl-th]} \BibitemShut {NoStop}%
\bibitem [{\citenamefont {Fromm}\ \emph {et~al.}(2012)\citenamefont {Fromm},
  \citenamefont {Langelage}, \citenamefont {Lottini},\ and\ \citenamefont
  {Philipsen}}]{Fromm:2011qi}%
  \BibitemOpen
  \bibfield  {author} {\bibinfo {author} {\bibfnamefont {M.}~\bibnamefont
  {Fromm}}, \bibinfo {author} {\bibfnamefont {J.}~\bibnamefont {Langelage}},
  \bibinfo {author} {\bibfnamefont {S.}~\bibnamefont {Lottini}}, \ and\
  \bibinfo {author} {\bibfnamefont {O.}~\bibnamefont {Philipsen}},\ }\href
  {\doibase 10.1007/JHEP01(2012)042} {\bibfield  {journal} {\bibinfo  {journal}
  {JHEP}\ }\textbf {\bibinfo {volume} {01}},\ \bibinfo {pages} {042} (\bibinfo
  {year} {2012})},\ \Eprint {http://arxiv.org/abs/1111.4953} {arXiv:1111.4953
  [hep-lat]} \BibitemShut {NoStop}%
\bibitem [{\citenamefont {Brown}\ \emph {et~al.}(1990)\citenamefont {Brown},
  \citenamefont {Butler}, \citenamefont {Chen}, \citenamefont {Christ},
  \citenamefont {Dong}, \citenamefont {Schaffer}, \citenamefont {Unger},\ and\
  \citenamefont {Vaccarino}}]{Brown:1990ev}%
  \BibitemOpen
  \bibfield  {author} {\bibinfo {author} {\bibfnamefont {F.~R.}\ \bibnamefont
  {Brown}}, \bibinfo {author} {\bibfnamefont {F.~P.}\ \bibnamefont {Butler}},
  \bibinfo {author} {\bibfnamefont {H.}~\bibnamefont {Chen}}, \bibinfo {author}
  {\bibfnamefont {N.~H.}\ \bibnamefont {Christ}}, \bibinfo {author}
  {\bibfnamefont {Z.-h.}\ \bibnamefont {Dong}}, \bibinfo {author}
  {\bibfnamefont {W.}~\bibnamefont {Schaffer}}, \bibinfo {author}
  {\bibfnamefont {L.~I.}\ \bibnamefont {Unger}}, \ and\ \bibinfo {author}
  {\bibfnamefont {A.}~\bibnamefont {Vaccarino}},\ }\href {\doibase
  10.1103/PhysRevLett.65.2491} {\bibfield  {journal} {\bibinfo  {journal}
  {Phys. Rev. Lett.}\ }\textbf {\bibinfo {volume} {65}},\ \bibinfo {pages}
  {2491} (\bibinfo {year} {1990})}\BibitemShut {NoStop}%
\end{thebibliography}%
  
\end{document}